\documentclass[twocolumn]{aastex6}
\usepackage{subcaption}
\usepackage{amsmath,amssymb,amstext}
\usepackage{xspace}
\usepackage{enumitem}

\newcommand{\Fermi}{\emph{Fermi}\xspace}
\newcommand{\AstroSat}{\emph{AstroSat}\xspace}

\newcommand*{\vcenteredhbox}[1]{\begingroup
\setbox0=\hbox{#1}\parbox{\wd0}{\box0}\endgroup}

\graphicspath{{figures/}}

\begin{document}

\title{Violation of synchrotron line of death by the highly polarized GRB~160802A}

\author{Vikas Chand}
\affiliation{Tata Institute of Fundamental Research,
Mumbai, India}
\author{Tanmoy Chattopadhyay}
\affil{Pennsylvania State University,
State College, PA, USA}
\author{S. Iyyani}
\affil{The Inter-University Centre for Astronomy and Astrophysics,
Pune, India}
\author{Rupal Basak}
\affiliation{The Oskar Klein Centre for Cosmoparticle Physics, AlbaNova, SE-106 91 Stockholm, Sweden; Department of Physics, KTH Royal Institute of Technology, AlbaNova University Center, SE-106 91 Stockholm, Sweden}
\author{Aarthy, E.}
\affil{Physical Research Laboratory,
Ahmedabad, Gujarat, India}
\author{A. R. Rao}
\affiliation{Tata Institute of Fundamental Research,
Mumbai, India}
\author{Santosh V. Vadawale}
\affil{Physical Research Laboratory,
Ahmedabad, Gujarat, India}
\author{Dipankar Bhattacharya}
\affiliation{The Inter-University Centre for Astronomy and Astrophysics,
Pune, India}
\author{V. B. Bhalerao}
\affiliation{Indian Institute of Technology, Bombay, India}

\begin{abstract}GRB~160802A is one of the brightest gamma-ray bursts (GRBs) observed with 
\textit{Fermi} Gamma-ray Burst Monitor (GBM) in the energy range of 10--1000\,keV,
while at the same time it is surprisingly faint at energies $\gtrsim2$\,MeV. An observation
with \textit{AstroSat}/CZT Imager (CZTI) also provides the polarisation which 
helps in constraining different prompt emission models using the novel joint
spectra-polarimetric data.
We analyze the \textit{Fermi}/GBM data, and find two main bursting episodes
that are clearly separated in time, one particularly faint in higher energies and having certain differences in their 
spectra. The spectrum in general shows a hard-to-soft evolution in both the episodes.
Only the later part of the first episode shows intensity tracking behaviour
corresponding to multiple pulses.   
The photon index of the spectrum is hard, and in over 90 per cent cases,
cross even the slow cooling limit ($\alpha=-2/3$) of an optically thin 
synchrotron shock model (SSM). Though such hard values are generally 
associated with a sub-dominant thermal emission, such a component is not statistically 
required in our analysis. In addition, the measured polarisation in 
100--300\,keV is too high, $\pi=85\pm29\%$, to be accommodated in 
such a scenario. Jitter radiation, which allows a much harder index up to
$\alpha=+0.5$, in principle can produce high polarisation but only beyond
the spectral peak, which in our case lies close to 200--300\,keV during 
the time when most of the polarisation signal is obtained. 
The spectro-polarimetric data seems to be consistent with a 
subphotospheric dissipation process occurring within a narrow 
jet with a sharp drop in emissivity beyond the jet edge, and 
viewed along its boundary.   
\end{abstract}


\section{Introduction}
  
One of the putative models invoked to explain the non-thermal spectral shape in the prompt emission of gamma-ray bursts
(GRB) is the synchrotron shock model (SSM).
In this model, electrons gyrating in the magnetic field at internal shocks generate synchrotron photons that are observed at 
gamma ray energies, boosted by the relativistic bulk motion of the jet (\citealt{Rees:1992}, \citealt{Meszaros:1993}, \citealt{Rees:1994}).
One of the predictions of the SSM is the so-called ``synchrotron line of death (LOD)''. The
low energy photon spectral index should not exceed the value $-2/3$ for an optically thin shocked material.
If the effects of the synchrotron cooling are also taken into consideration
(\citealt{Katz:1994}, \citealt{Sari:1996}, \citealt{Sari:1997}), then the index can 
lie in the range of -3/2 to -2/3. The distribution of the indices was, however, found to violate these
limits (\citealt{Cohen:1997}, \citealt{Crider:1997}, \citealt{Preece:1998}, \citealt{Ghirlanda:2003}).

The measurement of the low energy spectral index, however, depends on the spectral modelling 
of the GRB prompt emission. For most of the GRBs the shape of the spectrum could be 
phenomenologically well described by a Band function \citep{Band:1993}, which consists of two 
smoothly joined power law functions. The model parameters are low and high energy indices 
($\alpha$ and $\beta$), the energy where the $\nu F_\nu$ spectrum peaks ($E_p$) and
the normalization. Besides this empirical model, GRBs are known to show the 
evidence of other components in the
prompt emission spectrum. These include one or more thermal components 
modeled as a blackbody (\citealt{Ryde:2005}; \citealt{Page:2011}, \citealt{Guiriec:2011},
   \citealt{Guiriec:2013},  \citealt{Guiriec:2015b}, \citealt{BR:2015})
 or a non-thermal component modeled by a power law or cut-off power law extending up to
high energies ($>$100 MeV) that is observed in Fermi LAT energy band (\citealt{Gonzalez:2003}, \citealt{Abdo:2009}, \citealt{Ackermann:2013}).
In a unified model for prompt emission from optical to $\gamma - rays$, deviation from Band model is 
fit by a three-component model which includes two non-thermal components and a thermal component (\citealt{Guiriec:2015a},
 \citealt{Guiriec:2016a}, \citealt{Guiriec:2015b}).

Some studies show that the spectral evolution of a single emission component can 
sometimes make Band + BB artificially fit the data significantly better than Band alone if the integration time is too long.
Thus for an unambiguous detection of the thermal component, it is generally necessary to verify the presence 
of the thermal component in time resolved spectra as well \citep{Burgess:2015}.
Hence, though the inclusion of thermal component can alleviate the problem of line of death (LOD)
violation, we must be cautious of such cases. In addition,
\citet{Burgess:2014} have  used a physical synchrotron model with a blackbody
and found that 
the problem of LOD violation persists in many cases.
They have also found a more severe LOD $\alpha \sim -0.8$ and pointed out the
need for some other emission mechanisms. 
One such proposed mechanism is Jitter radiation. These radiations are
emitted by ultra relativistic electrons in a non-uniform, small-scale magnetic field and 
produces a spectral shape that is different
from the synchrotron radiation \citep{Medvedev:2000}. The allowed photon index 
in a Jitter radiation can reach up to +0.5.

From data analysis point of view, GRBs with high signal to noise spectral data 
are ideal to try out different emission models as the prompt emission shows rapid spectral 
evolution and these GRBs provide good enough signal for time resolved studies. 
However, the spectral and timing data so far have not been able to pin down the 
radiation mechanism. It is thus very important to study bright GRBs with well 
defined spectral shapes using other informations such as X-ray polarisation 
during the prompt emission phase. As different emission models predict different 
degree of polarisation in different energy bands, this is a powerful technique 
to provide strong constraints on the possible models.
GRB~160802A is a bright GRB showing significant hard X-ray polarization \citep{Chattopadhyay:2017}. It shows two pulses 
in its lightcurve and the spectrum is well fit by a simple Band function and hence it offers 
a very good opportunity to carry out a simultaneous timing, spectral and polarisation
study of a class of GRBs violating LOD.

In Section \ref{sec:Fermi_Astrosat} we discuss the joint usage of Fermi and $AstroSat$ for
spectral and polarization studies of GRBs. Then we present the timing and spectral properties 
of the GRB in Section \ref{sec:160802A}.
We conclude and discuss our results in Section \ref{sec:conclusions_discussions}.

\section{Fermi and AstroSat data}
\label{sec:Fermi_Astrosat}
For many years, $Neil$ $Gehrels$ $Swift$ $Observatory$ and $Fermi$ satellites have been providing detailed information on the prompt 
emission of GRBs \citep{Gehrels:2013}. The Burst Alert Telescope (BAT) on-board Swift \citep{Gehrels:2004} is a 
dedicated instrument to detect GRBs and the satellite slews and points towards the location of a 
GRB during the prompt emission.
BAT, however, has a relatively narrow energy band and hence the spectrum of the prompt emission for the 
BAT detected GRBs can be generally modeled as a simple powerlaw.
Gamma-Ray Burst Monitor (GBM) on board \Fermi is comprised of 12 sodium iodide (NaI) 
detectors and 2 bismuth germanate (BGO)
detectors  \citep{Meegan:2009a}. These detectors are sensitive
in the 8 keV $-$ 1 MeV and 150 keV $-$ 40 MeV energy range, respectively. 
In addition,
the  Large Area Telescope (LAT) on board \Fermi, is sensitive from 20 MeV to 300 GeV \citep{Atwood:2009}.
The unprecedented coverage over seven decades in energy by \Fermi has  led to the discovery of 
substantially new science for GRBs. The \Fermi/LAT GRB catalog contains several interesting bright bursts and \Fermi helped revealing a new spectral component 
that exists up to GeV energies or spectral breaks existing in the MeV energy ranges (\citealt{Abdo:2009}, \citealt{Izzo:2012}, \citealt{Ackermann:2010}, 
\citealt{Vianello:2017}, \citealt{Wang:2017}). Still it is 
generally felt that the spectral modelling alone is unable to 
solve the problem of radiation mechanism of the GRB prompt emission due to various issues. Some key problems are e.g.,
(a) the same data can be fit with a variety of models and the true 
model cannot be determined based on the goodness of fit; (b) even if the 
best-fit model is determined, the models are generally phenomenological and 
may not conform with the underlying theory e.g., the LOD violation; (c) sometimes 
an additional spectral component e.g., a blackbody may change the parameters of 
the other component e.g., Band function in such a way that it conforms with the underlying model, 
but the additional component is not statistically required. 
A critical component that can break these degeneracies inherent in
spectral modelling is the measurement of X-ray polarisation. The Cadmium
Zinc Telluride Imager (CZTI) on board AstroSat is highly sensitive to hard X-ray polarisation (\citealt{Chattopadhyay:2014}, 
\citealt{Vadawale:2015}, see also Section 4 of \citealt{Chattopadhyay:2017}).

\AstroSat is a multi-wavelength observatory which was
launched on 2015 September 28 \citep{Singh:2014}. The CZTI focal plane consists of pixellated detectors
sensitive in the energy range of 20-200~keV, with sensitivity gradually
falling off till about 500~keV. All CZTI data are acquired in ``event
mode'', with individual photons time-tagged at 20~$\mu$s resolution \citep{Bhalerao:2017}.
CZTI can help the study of GRB prompt emission by measuring
X-ray polarization in the 100 - 300 keV range (\citealt{Chattopadhyay:2014}; \citealt{Vadawale:2015}).
The photons preferentially scattered in the direction perpendicular to the polarization direction,
give rise to an asymmetry/modulation in an otherwise flat azimuthal angle distribution.
Amplitude of the modulation is directly
proportional to the polarization fraction embedded in the incident radiation.
Selection procedure of the Compton events in CZTI is discussed in detail in
\citet{Chattopadhyay:2014}.

The spectral and temporal properties from \Fermi  and polarization from the
$AstroSat$/CZTI can give a complete information about a GRB. We need to study these gathered information individually as well as a statistical sample of it. 
GRB~160802A is one of the GRBs observed in both \Fermi and $AstroSat$/CZTI.
We present here a combined temporal, spectral, and polarization characteristics of this GRB.

\section{GRB~160802A}
\label{sec:160802A}

\subsection{Observations}

\begin{figure*}
\centering
\vcenteredhbox{\includegraphics[width=0.49\textwidth]{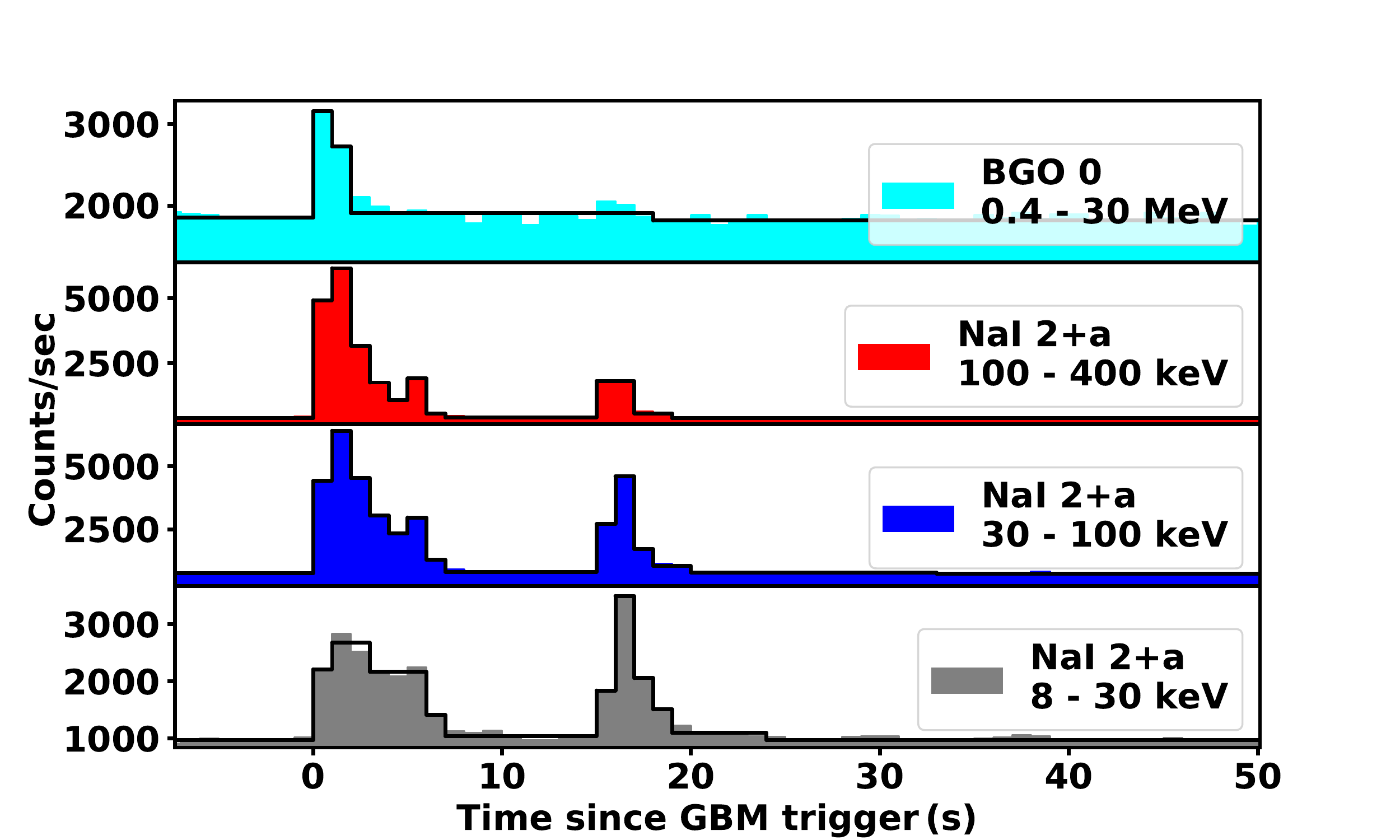}}
\vcenteredhbox{\includegraphics[width=0.49\textwidth]{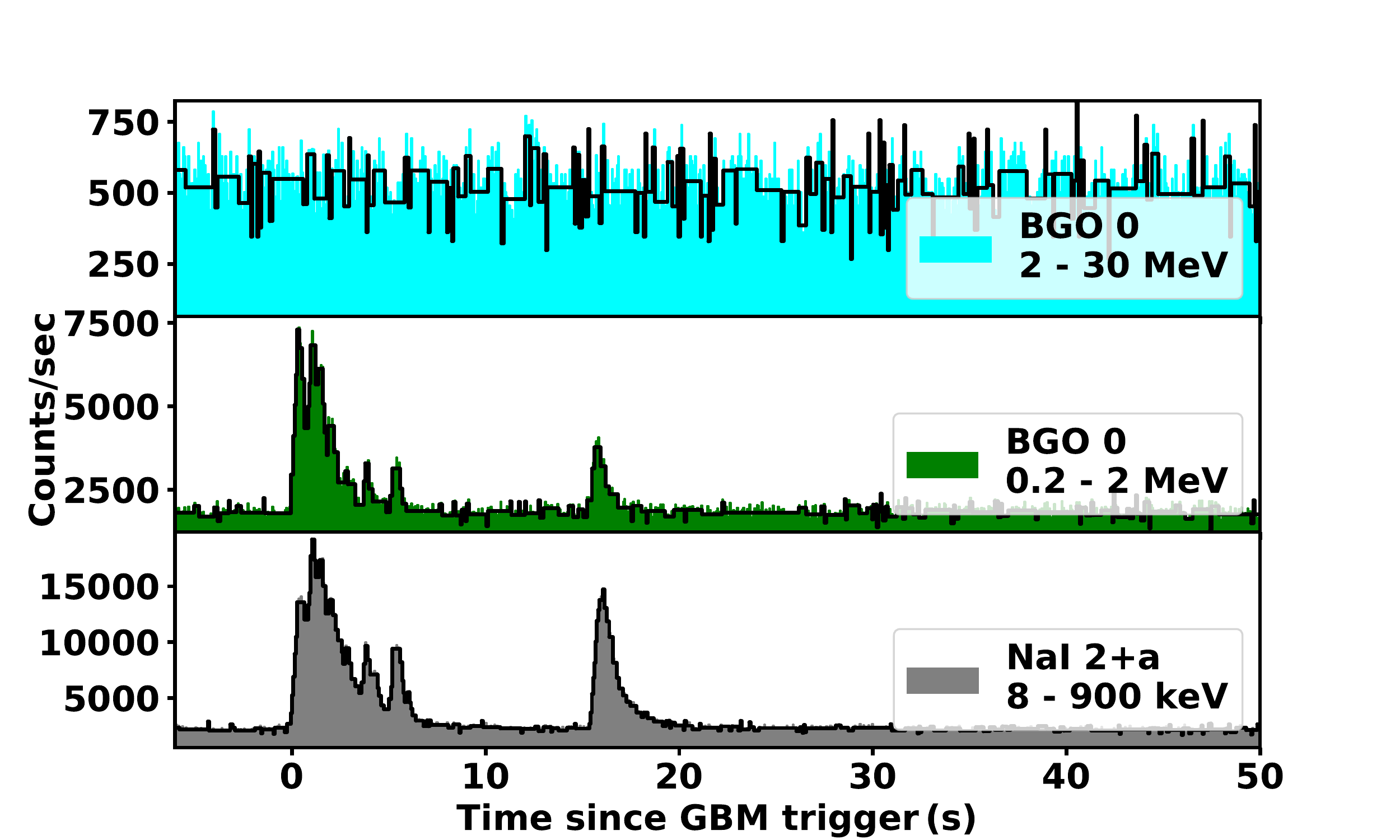}}
\caption{Left: Composite light curve with 1 s resolution showing cumulative
	rates from two NaI detectors (NaI 2, a) and BGO 0. Bayesian blocks are
	overplotted and show the light curves obtained using these blocks of
	constant rate. Right: light curve with a higher temporal resolution of
	64 ms.  The emission is limited up to energy $\sim$ 2 MeV in the BGO as
	shown in the top panel.}
\label{fig:160802A_LC}
\end{figure*}

GRB~160802A was detected by $Fermi$/GBM at UT~06:13:29.63 \citep{Bissaldi:2016}
as a very bright GRB with two peaks, and a $T_{90}$ of $16.4\pm0.4$ s in the
50--300~keV band (see \citealt{Kouveliotou:1993} for a definition of T$_{90}$). Both the peaks showed a Fast Rise Exponential Decay
(FRED)-like profile.  The peak energy of the Band function fit to the
time--integrated spectrum ($T_0-0.3$~s to $T_0+19.4$~s) is $284 \pm 7$~keV in
the preliminary analysis.  The fluence as observed in the 10-1000 keV band is
$1.04\pm0.08 \times 10^{-4}$ erg~cm$^{-2}$.  $AstroSat$/CZTI data show a complex
light curve with $T_{90}$ = 16.8~s \citep{Bhalerao:2016}.
The burst was also
detected by several other satellites including the {\it Block for X-ray and gamma-radiation detection Lomonosov,~BDRG}
(\citealt{Panasyuk:2016}), Wind/Konus (\citealt{kozlova:2016}) and the {\it Calorimetric Electron Telescope, CALET}
(\citealt{tamura:2016}). No low energy (X-rays, optical, radio etc.) and high energy ($GeV$) afterglows were 
reported for GRB~160802A\footnote{\url{https://gcn.gsfc.nasa.gov/other/160802A.gcn3}}.\\

\subsection{Light curves}
\label{sec:lightcurves}
Lightcurves obtained from NaI and BGO detectors of \emph{Fermi}-GBM clearly show two peaks 
separated by $\sim8$~s in GRB~160802A (Fig.~\ref{fig:160802A_LC}). The left panels show 
lightcurves in different energy bins, plotted at 1~second resolution.
Bayesian blocks \citep{Scargle:2013} are obtained using $Fermi$ $science ~tool$ $gtburstfit$ with $ncpprior$ parameter set to 9.
A fit obtained by 
Bayesian blocks analysis is overplotted with a solid black line. We see that the second episode 
is softer: with a high count rate at low energies, but rapidly diminishing above 400~keV. 
We studied lightcurves with finer energy bands to identify bands with significant emission, and 
found that at energies $\gtrsim 2$~MeV, any putative signal in BGO is indistinguishable from 
background. To highlight these features, the right panel of Fig.~\ref{fig:160802A_LC} 
shows lightcurves with 64~ms time resolution.

\subsection{Spectral analysis of the prompt emission}
\label{sec:spectral_analysis}
We undertake detailed spectral analysis with data from GBM detectors with the
strongest signal: NaI~2 (n2), with the GRB just $20\,^{\circ}$ from its
boresight, and NaI~a (na), which detected the GRB at an off-axis angle of
$54\,^{\circ}$. We also select the BGO detector closest to the GRB direction
(BGO~0). We use \textit{Time-Tagged-Events} (TTE) data provided by the GBM team
and publicly available on the FSSC\footnote{Fermi Science Support Center,
\url{https://fermi.gsfc.nasa.gov/ssc/}.} website. We generate custom response
matrices and spectrum files using the public software
\texttt{gtburst}\footnote{\url{https://fermi.gsfc.nasa.gov/ssc/data/analysis/scitools/gtburst.html}}
and using best localization available interplanetary network (IPN) triangulation \citep{Kozlova:2016b}. 
The response file is a weighted response for an interval split 
over multiple extensions where each extension contains response for a particular time interval 
which can vary within extensions.
The software $gtburst$ assign weights when a selected time-bin for spectral analysis is split across two or multiple extensions.
The spectra were analyzed
in $XSPEC$ \citep{Arnaud:1996}. 
Spectra reduced using $gtburst$ are PHA 2 files and can be directly used in $XSPEC$ by referencing the spectrum number.
Our data are Poissionian in nature, with a Gaussian
background derived from modelling the spectrum in intervals before and after the
GRB emission. Hence, we use
\texttt{pgstat}\footnote{\url{https://heasarc.gsfc.nasa.gov/xanadu/xspec/manual/node293.html}}
as the data fit estimator. Effective area corrections were applied among the GBM and BGO detectors
and to avoid the k-edge 33-37 keV energy range was excluded for the analysis of NaI data.
There is a trade-off between the reduction in
$pgstat$ and increase in the number of model parameters used to fit the data.
This comparison can be made by using the Bayesian Information Criterion (BIC).
The BIC can be calculated from the $pgstat$ value, the number of free parameters
to be estimated ($k$), and the number of data points ($n$) as $$\mathrm{BIC} =
-2 \ln(L) + k \ln(n)$$ where $L$ is the likelihood of a model for best fit
parameters \citep{Schwarz:1978}. When written in terms of
likelihood function we can have $pgstat = -2\ln(L)$, where $L$ is the
likelihood of a model for best fit parameters \citep{Schwarz:1978}. A model
yielding a lower BIC as compared to other models is taken to be preferred model,
based on the magnitude of the reduction in BIC.  A change of  $\Delta BIC\ge6$
is a strong evidence of improvement \citep{Kass:1995}.  An example of using BIC
for distinguishing among different models used to fit the GRB spectrum can be
found for the case of GRB~160625B in \cite{Wang:2017}.

We find that for GRB~160802A, the time-integrated \emph{Fermi} spectrum is best
fit by a Band function with $E_p$ = $276_{-14}^{+15}$ keV, $\alpha$ =
$-0.71_{-0.03}^{+0.03}$ and $\beta$ = $-2.5_{-0.2}^{+0.1}$.  Here $\alpha$
satisfies the LOD condition, $\alpha< -2/3$.  A thermal blackbody component
added to the spectrum improves the fit: $\Delta
pgstat \sim 16 ~(\Delta BIC = 3)$ for 2 more free parameters, although this is only a hint for the presence
of a thermal component and a detailed spectral analysis is further required.

\begin{figure*}
\centering
\includegraphics[scale=0.325,angle=270]{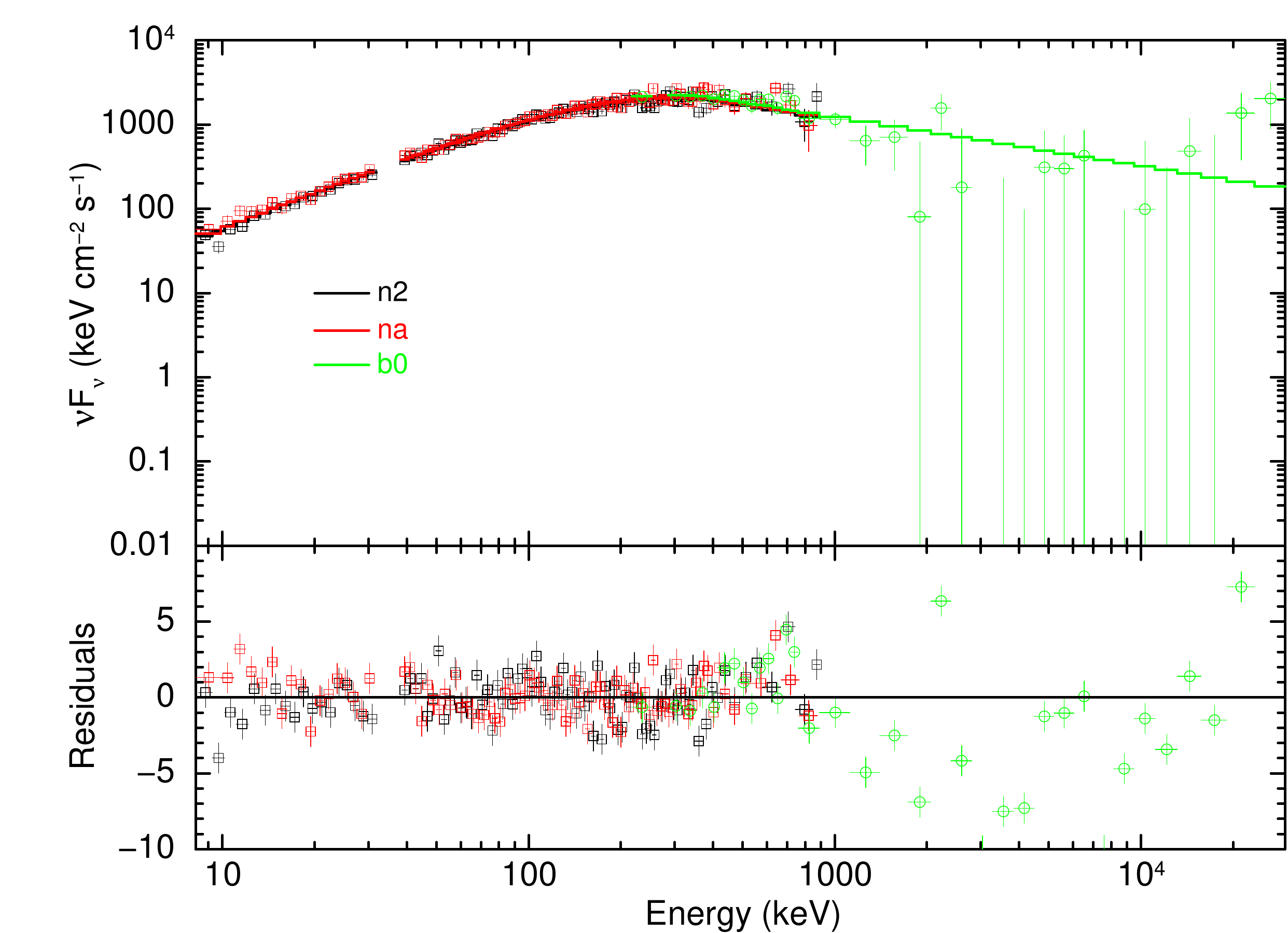}
\includegraphics[scale=0.325,angle=270]{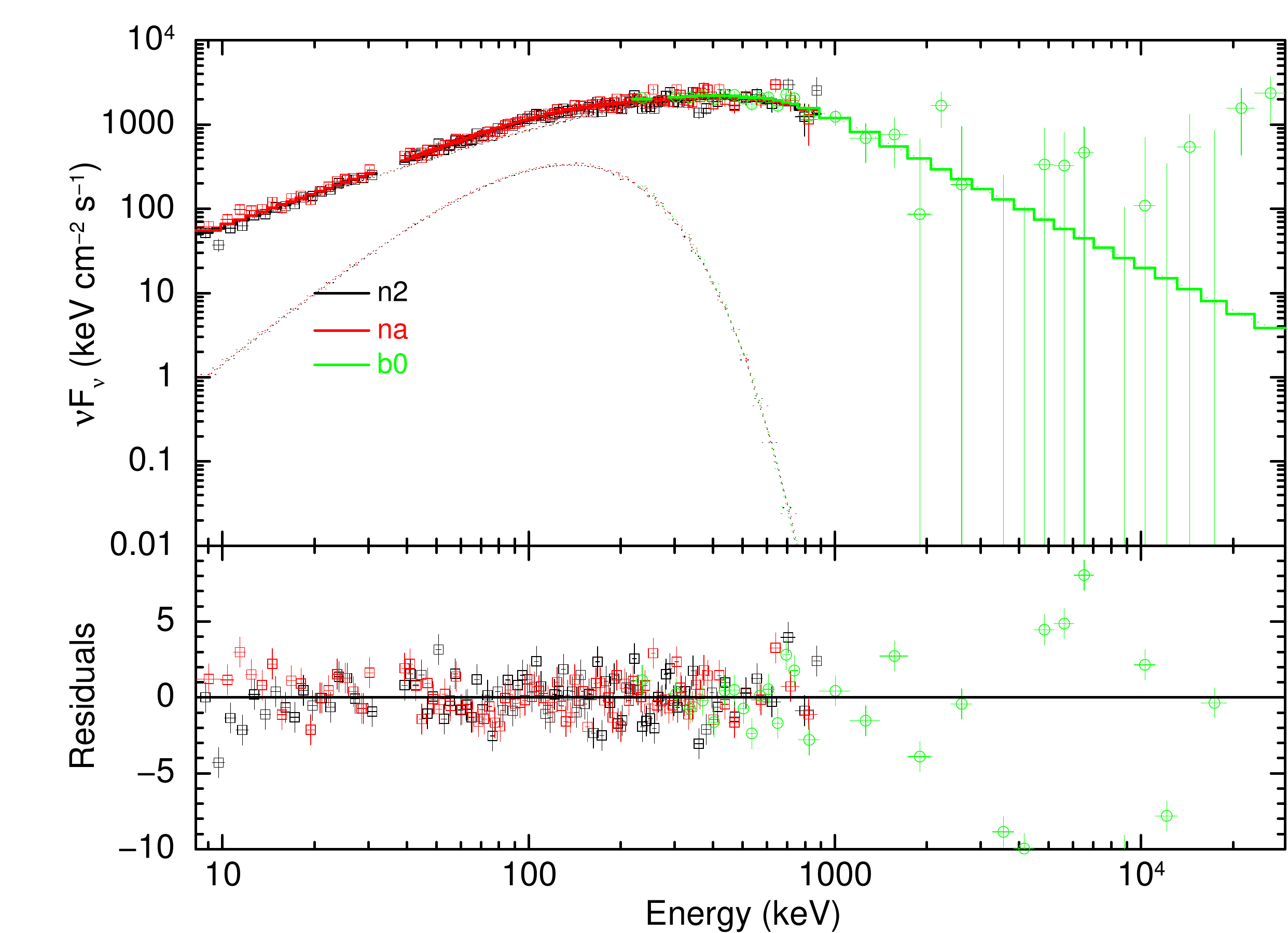}

\includegraphics[scale=0.325,angle=270]{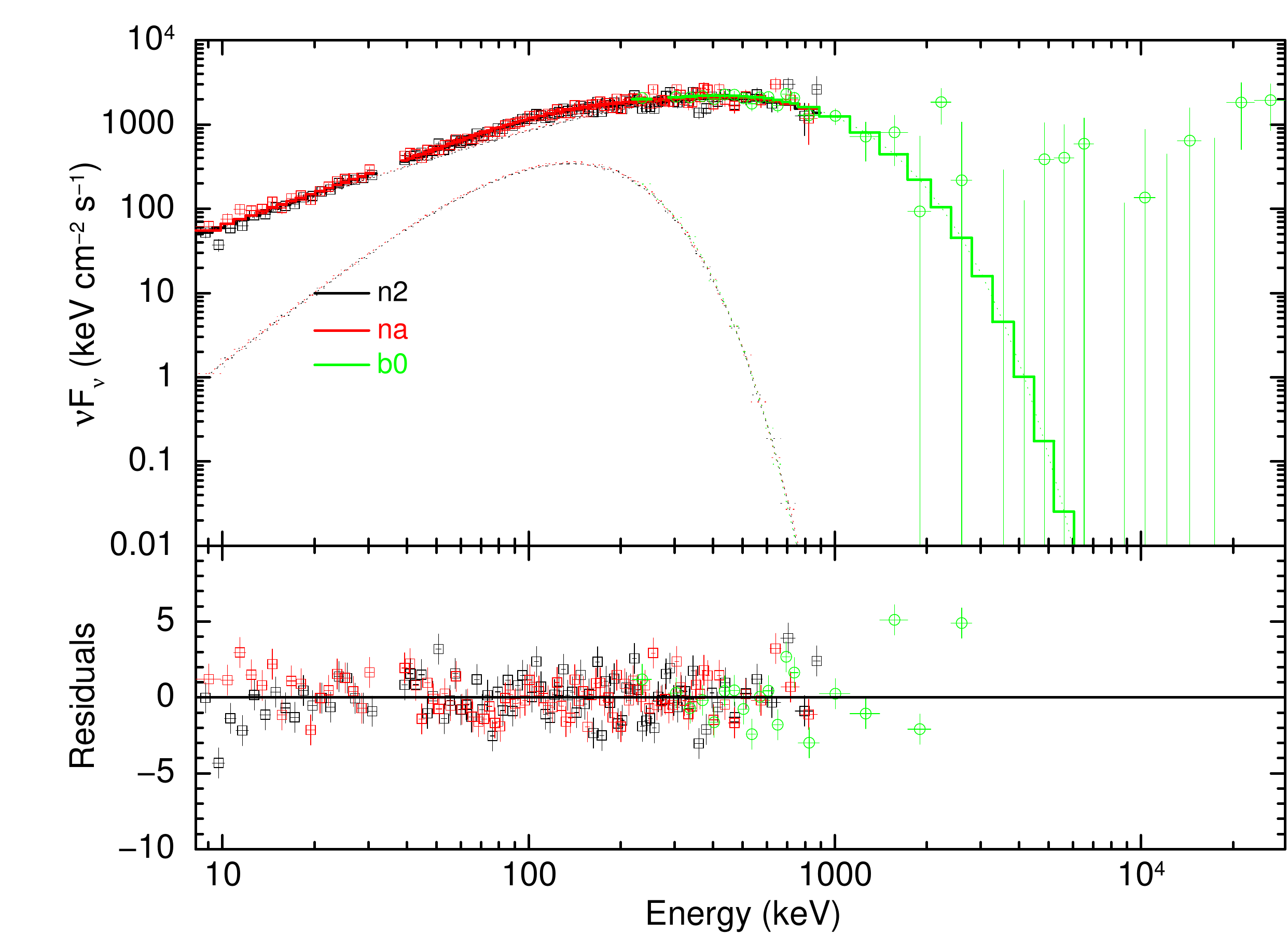}
\includegraphics[scale=0.325,angle=270]{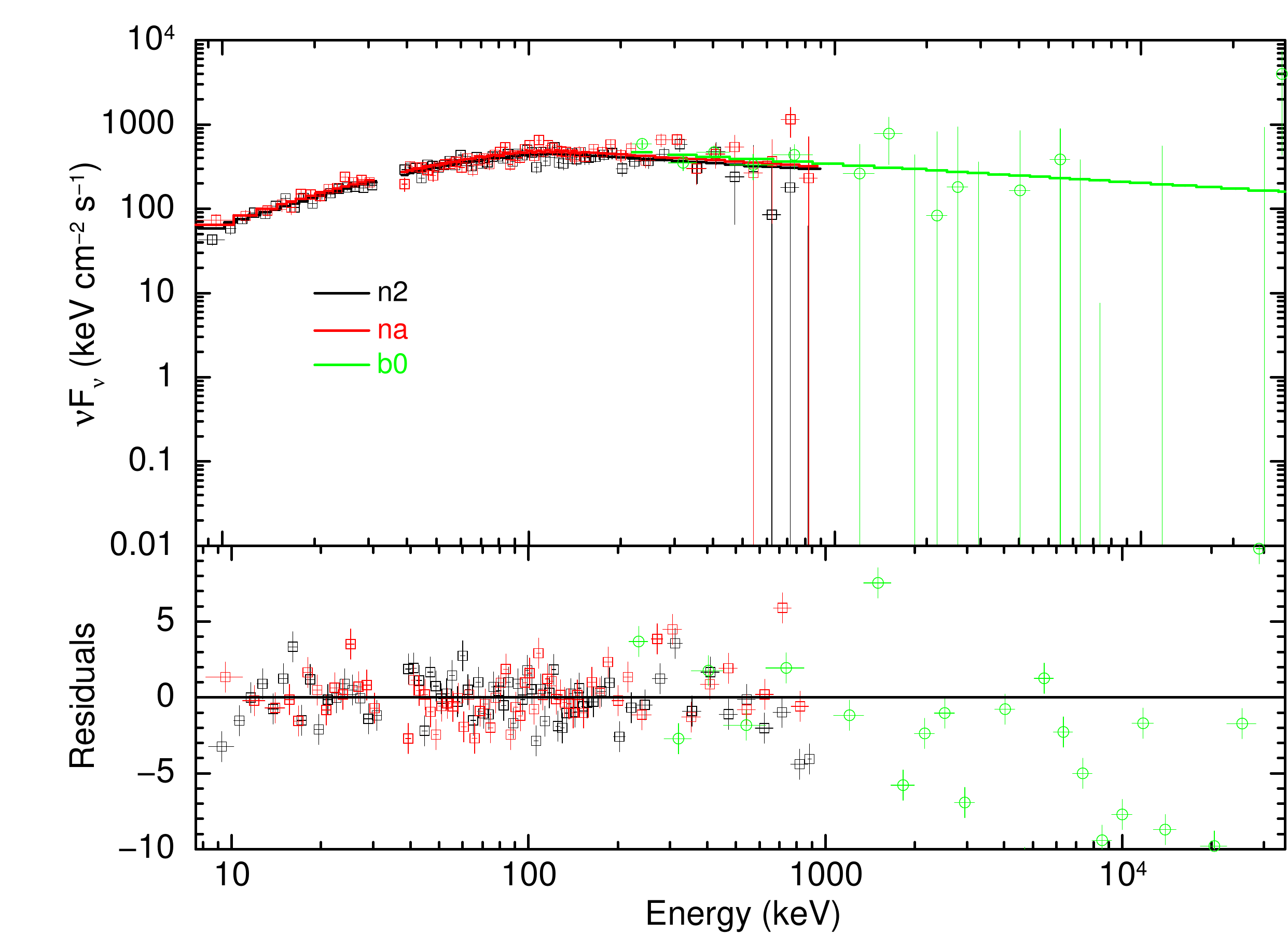}
\caption{The unfolded time integrated spectrum of GRB~160802A for pulse 1 \& 2 are shown along with the residuals (limited between -10 to 10) to
the fit. The  Band model (top left), an additional blackbody (top right) and blackbody added to a powerlaw with an exponential cut-off (bottom left)
are for the pulse 1 and a band function fit (bottom right) is shown for pulse 2.}
\label{fig:pulse_1_2_spectra}
\end{figure*}

Then we perform a spectral fitting to the two episodes separately using Band and
cut-off power-law (CPL) models, with or without a blackbody (BB) component and
the results are given in Table~\ref{tab:spec_gbm_lat_3GeV}.  We find that for
the first episode $\alpha$ of Band-only fit is $-0.52_{-0.03}^{+0.04}$, violating the
synchrotron LOD for slow cooling. Addition of a blackbody to the Band function
softens the value of $\alpha$ to $-0.73_{-0.05}^{+0.06}$ which is now consistent with
the LOD condition. The addition of the blackbody to either Band or CPL model
significantly improves the fit for the first episode with $\Delta BIC=16$ and
$55$, respectively. For the second episode the addition of the blackbody
component is required only for the CPL model ($\Delta BIC=24$), while indeed it
is disfavoured in case of Band function with $\Delta BIC=-6$. The spectra for time-integrated analysis are shown in Figure \ref{fig:pulse_1_2_spectra}.  A blackbody in the
integrated spectrum is just a first step towards finding a thermal component.
In order to verify that the added blackbody component is physical and not an
artefact of an evolving Band function, we then resort to time resolved spectral
analysis.

\subsection{Time resolved analysis with coarse bins}
\label{sec:spectral_analysis_tr_coarse}
\begin{figure*}
\centering
\includegraphics[scale=0.325,angle=270]{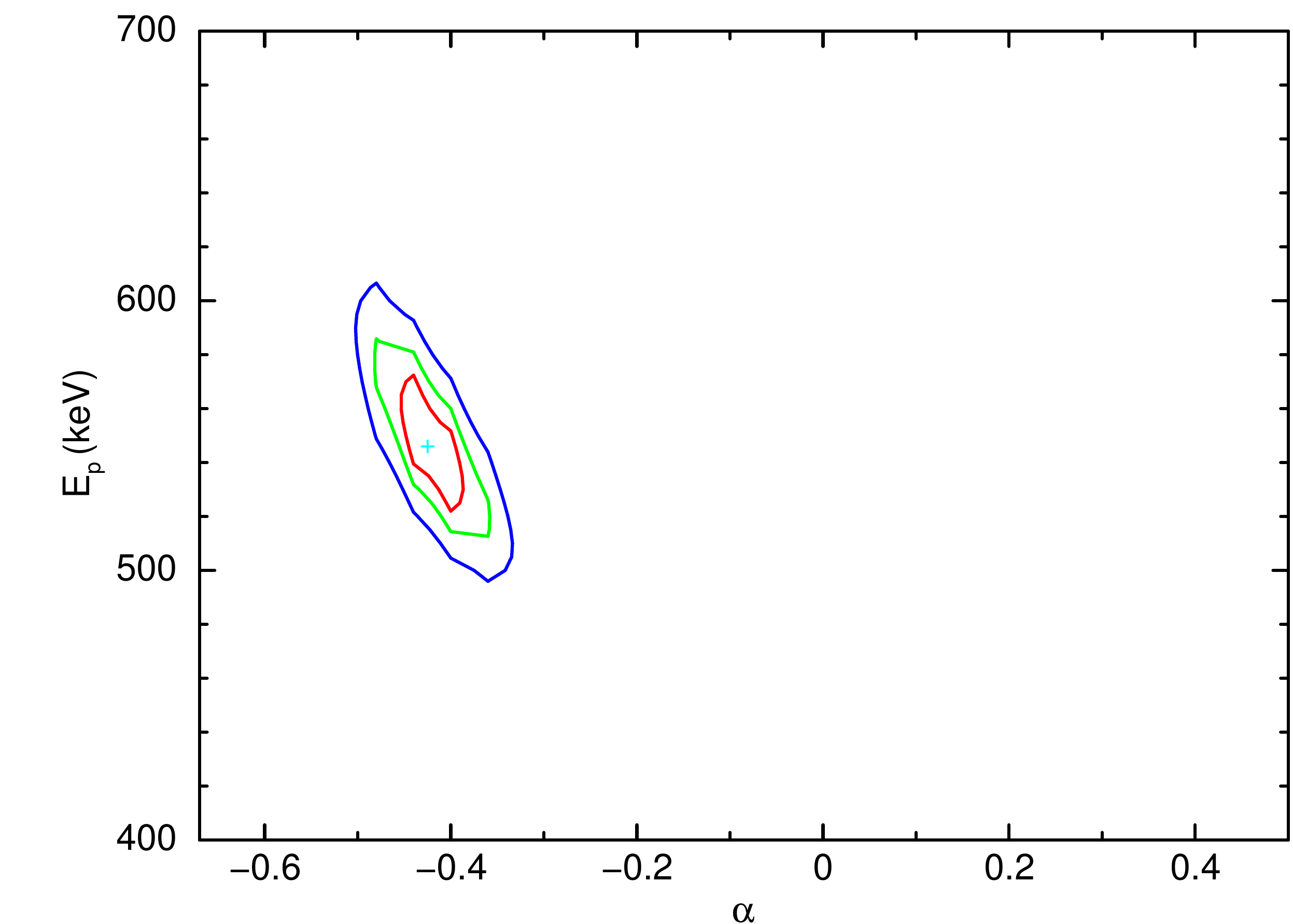}
\includegraphics[scale=0.325,angle=270]{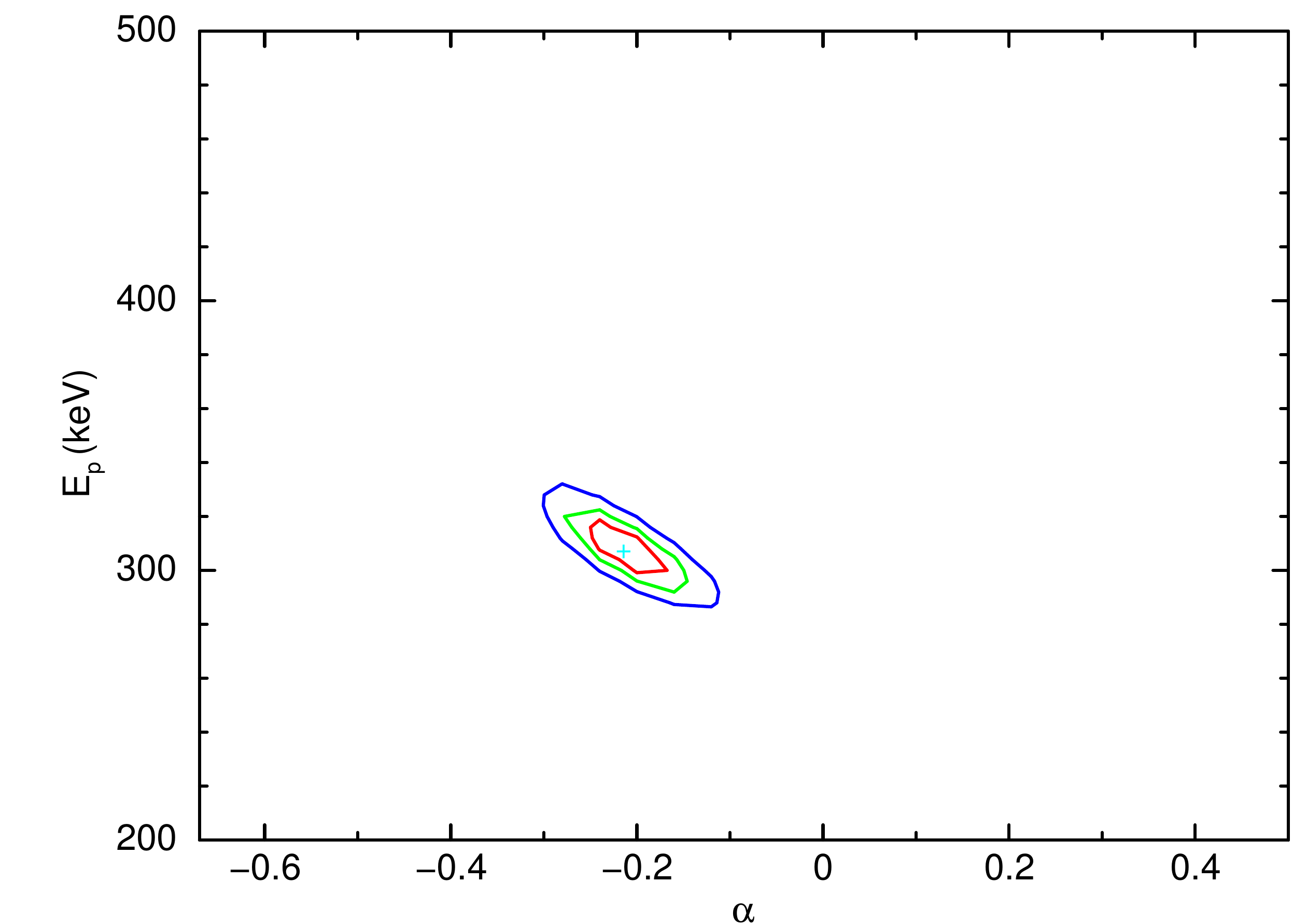}
\includegraphics[scale=0.325,angle=270]{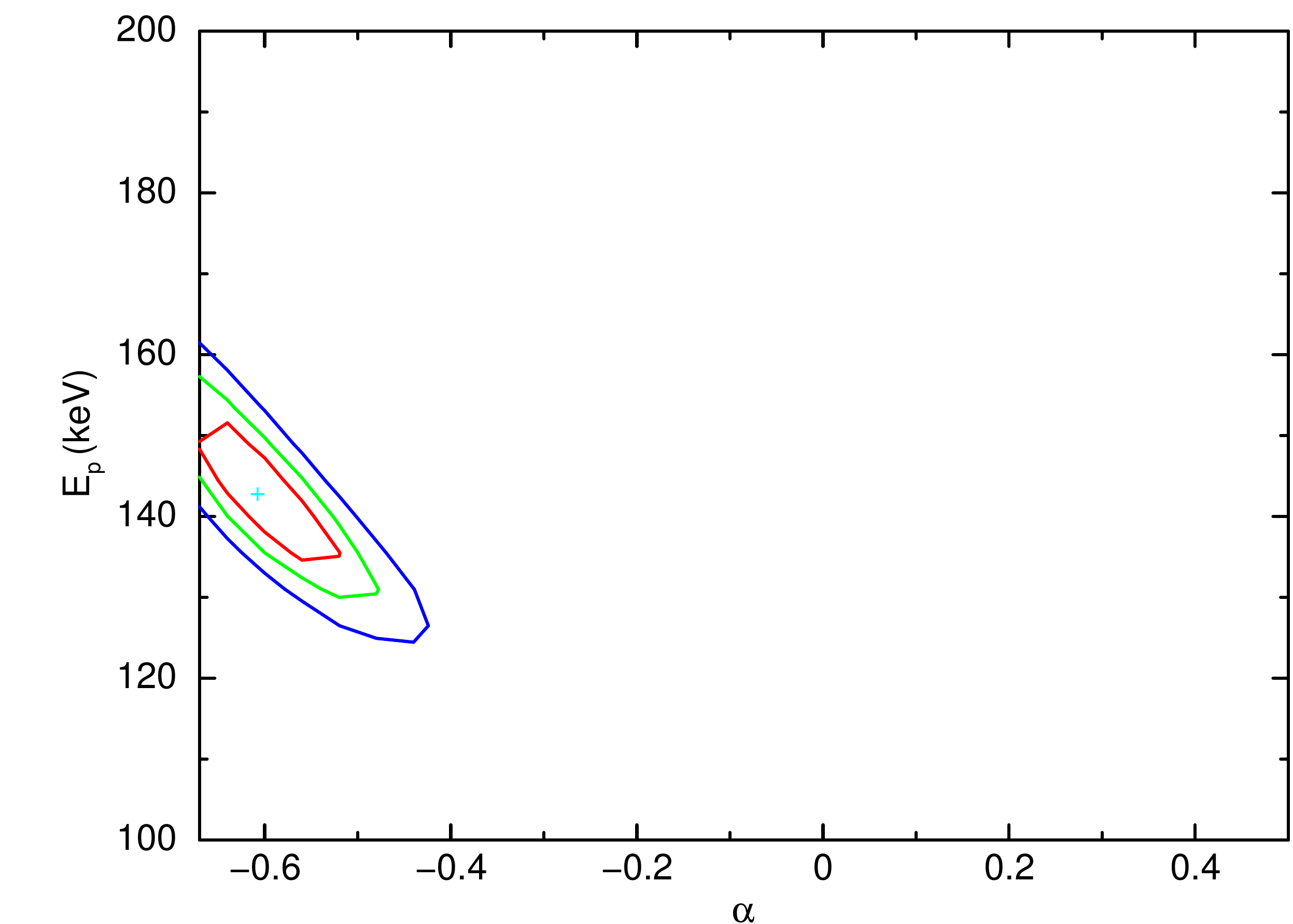}
\includegraphics[scale=0.325,angle=270]{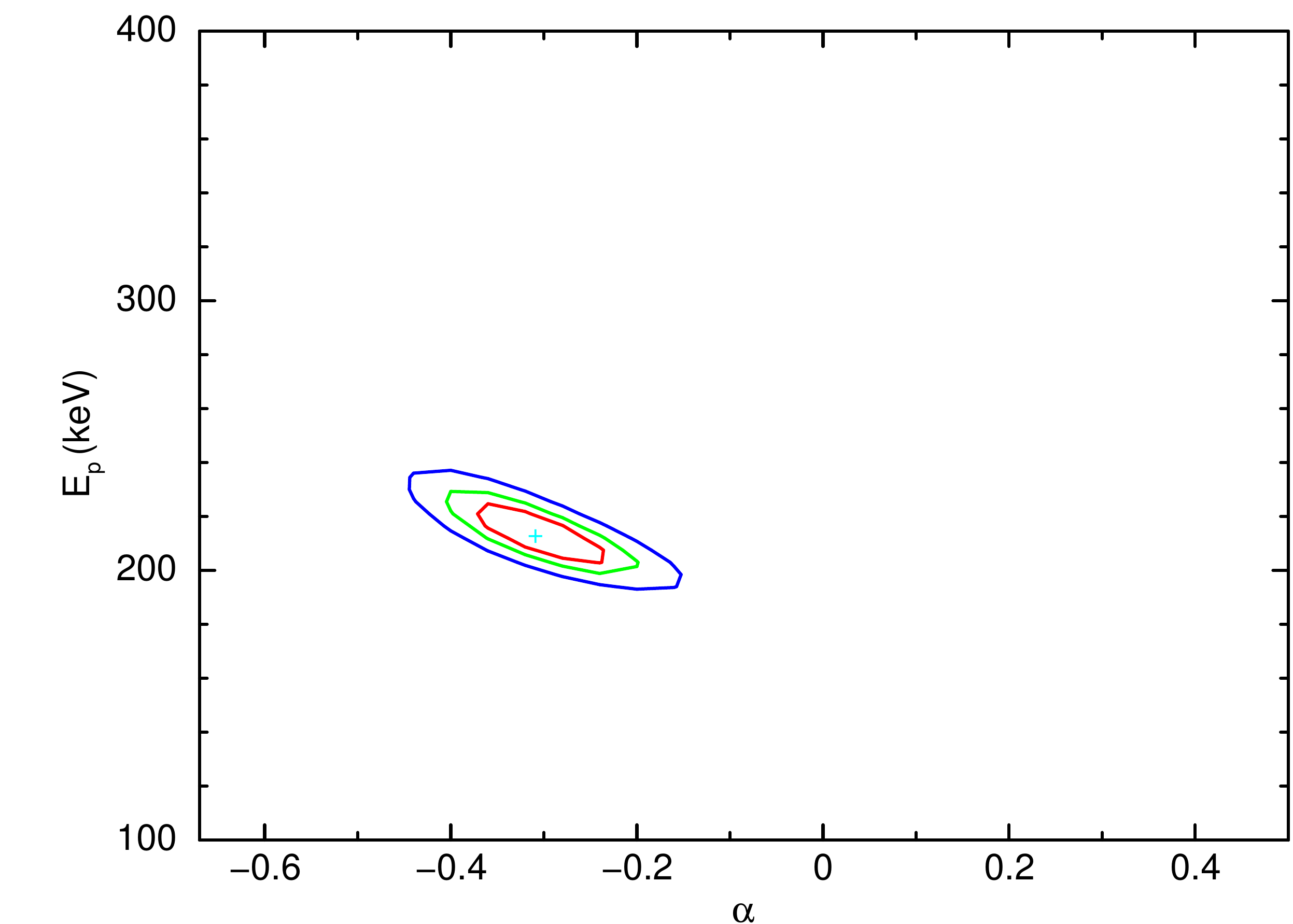}
\caption{The confidence contours of $\alpha$ and $E_{p}$ are shown for the
	coarse time bins (i), (ii), (iii) and (viii) clockwise from the top
	left. For the first three intervals which spans time interval $0~-~ 3~
	s$, the value of $\alpha$ is greater than what is expected from the
	popular SSM. For (viii) however, the $\alpha$ encompasses the allowed
	value in its $2\sigma$ contour. Y-axis ($E_p$) is drawn at the line of death value $-2/3$.}
\label{fig:alpha_contours}
\end{figure*}

We divide the GRB lightcurve into coarse bins corresponding to the Bayesian
blocks (BB) obtained from 1 second light-curves (\S\ref{sec:lightcurves}). The
Bayesian blocks algorithm objectively divides the data into an optimum set of
blocks with no statistically significant variation from a constant rate
within each block
\citep{Scargle:2013}. The time intervals thus obtained along with the models
tested are given in Table~\ref{tab:coarse_results}.  The synchrotron LOD is
violated in the first three intervals as is evident by the contour maps of
$\alpha$-$E_p$ shown in Figure~\ref{fig:alpha_contours}. In the
intervals (i), (ii) \& (viii), BB added to Band gives equally well fit and for intervals (iv), (vii) and (ix), a powerlaw with sharp break
($XSPEC$ model $bknpower$) is at par with Band function. With the addition of a BB, the value of the spectral index
$\alpha$ softens in all the time bins. The presence of BB can also be an
artefact of the evolution of Band function parameters with time.  This can be
tested by fitting spectra to smaller time bins, and requiring the smooth
evolution of parameters like the blackbody temperature with time \citep[see for
instance][]{Guiriec:2011, Guiriec:2013, Burgess:2015}. Motivated by the results from coarse analysis, we
explore this further in \S\ref{sec:spectral_analysis_tr_detailed}.

\subsection{Detailed time-resolved analysis}
\label{sec:spectral_analysis_tr_detailed}
The time intervals are obtained by choosing a constant signal to noise ratio
(SNR) of 30 in the light curve obtained from NaI detector $n2$ (which has the
highest observed rates). We found that the addition of a blackbody does not give
a smooth evolution of its parameters, the temperature of BB fluctuates
erratically and in many bins, it remained unconstrained. Moreover the Band
function emerges as the preferred model over BB+Band in all the bins, as
evidenced by the increase in BIC values after adding the blackbody component
(see Table~\ref{tab:specfitting}). We conclude that we do not detect any
significant blackbody component in the prompt emission of GRB~160802A.

\begin{figure*}
\centering
\includegraphics[scale=0.5,angle=0]{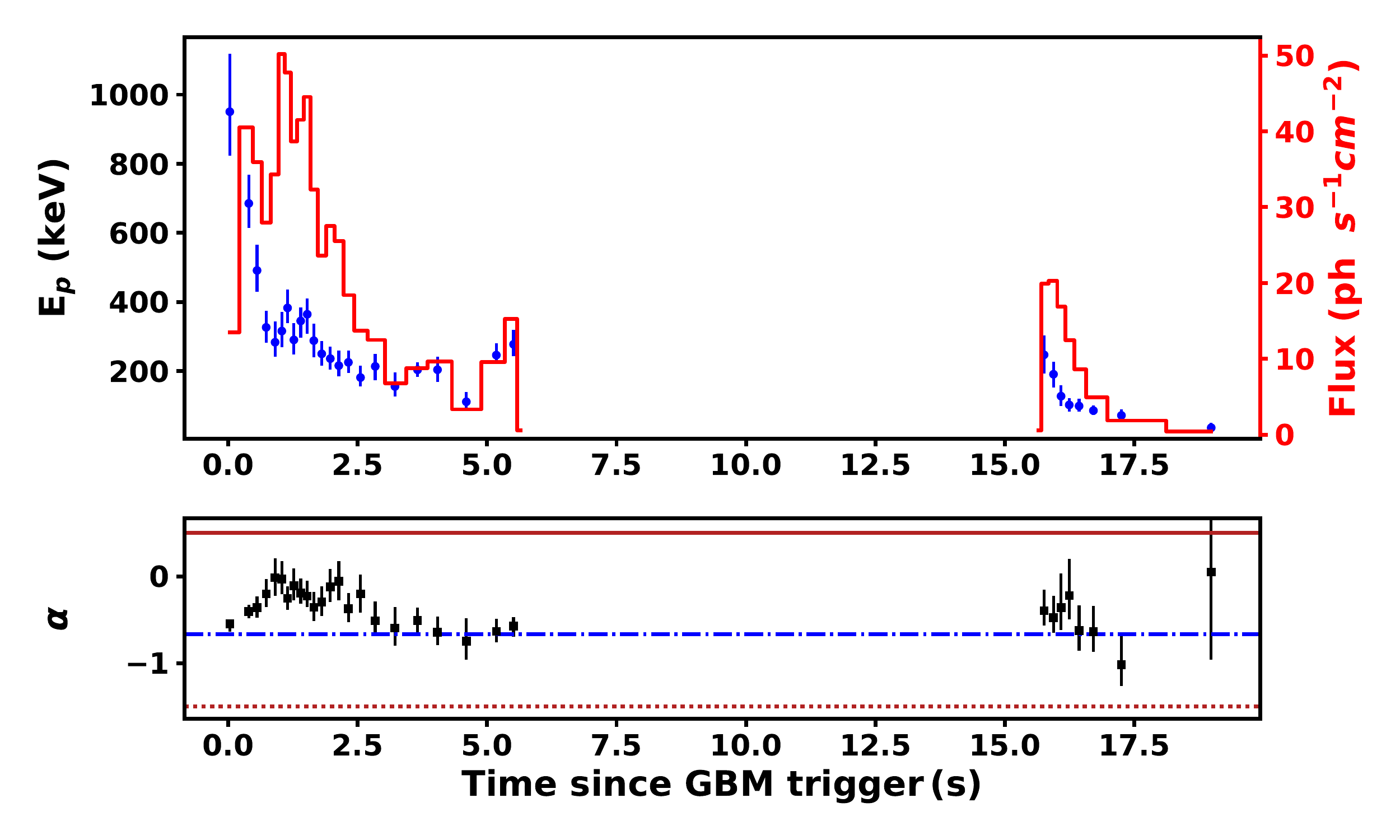}
\caption{The $E_p$ evolves with time like a hard to soft (HTS) evolution
	initially and then intensity tracking for the first episode. Second
	episode has HTS evolution of its $E_p$. The photon flux calculated in 8 - 1000 keV is also shown in
	the background.  The three horizontal lines in the second panel are
	lines of death for synchrotron fast cooling ($\alpha = -3/2$, dotted red
	line), synchrotron slow cooling ($\alpha = -2/3$, blue dash-dot line),
	and Jitter radiation ($\alpha = +1/2$, solid red line).}
\label{fig:160802A_alpha_evolution}
\end{figure*}

The evolution of the peak energy ($E_p$) of GRBs has been found to show hard to
soft evolution with time in some GRBs, $E_p$ tracks the intensity variations in
others, while some GRBs show a mixture of both phenomena \citep{Lu:2012}. An
exception to this behavior is found in the case of GRB~151006A
\citep{Basak:2017} where a hardening (increase) in $E_p$ is found in an
apparently single pulse towards the end of the emission. The variation of the
spectral parameters along with flux (Figure \ref{fig:160802A_alpha_evolution})
show the $E_p$ values for the first episode initially follow a hard to soft
(HTS) trend, and then track the 8--900~keV flux. In particular, the evolution of
$E_p$ is very fast in the first second: showing an order of magnitude change as
it decreases from $\sim 1000$~keV to 200~keV.  The fast evolution is likely to
have manifested itself as a curvature in the spectrum, which was incorrectly
modeled as a blackbody in the coarse analysis
(\S\ref{sec:spectral_analysis_tr_coarse}).

From the evolution of $\alpha$ we find that the values are crossing the LOD as
shown in Figure \ref{fig:160802A_alpha_evolution}.  The results are unaltered if
we use only one of the detectors $n2$ that has pointing angle $<50^{\circ}$.
The value of $\alpha$ and $E_p$ are determined by the data below the break
energy, while the $\beta$ is determined from data at the energies above that.
Although Fermi GBM covers the whole of the burst, the number of photons are very
few at energies in the BGO band. Therefore $\beta$ remains unconstrained in most
of the bins and is manifested as a steep high energy power law component of the
Band function (in $\nu F_\nu$ representation) because of large magnitudes of the
nominal best-fit values.

\subsection{Polarization analysis}
\label{sec:polarization}

\citet{Chattopadhyay:2017} made a systematic analysis of the GRBs detected by
CZTI during the first year of its operation and have reported positive
polarisation detection (chance probability $<$0.1) for 5 of the 11 GRBs having
sufficient number of Compton events for polarisation analysis. GRB~160802A shows
a high degree of polarisation ($85 \pm 30 ~\%$) and it has the second highest
Bayes factor for the polarized model as compared  to the unpolarized model among
all the GRBs. We have explored the polarisation characteristics of this GRB with
an aim to optimise the polarisation measurement in terms of the selected energy
and time windows.

\begin{figure} 
	\centering \includegraphics[scale=0.68,angle=0]{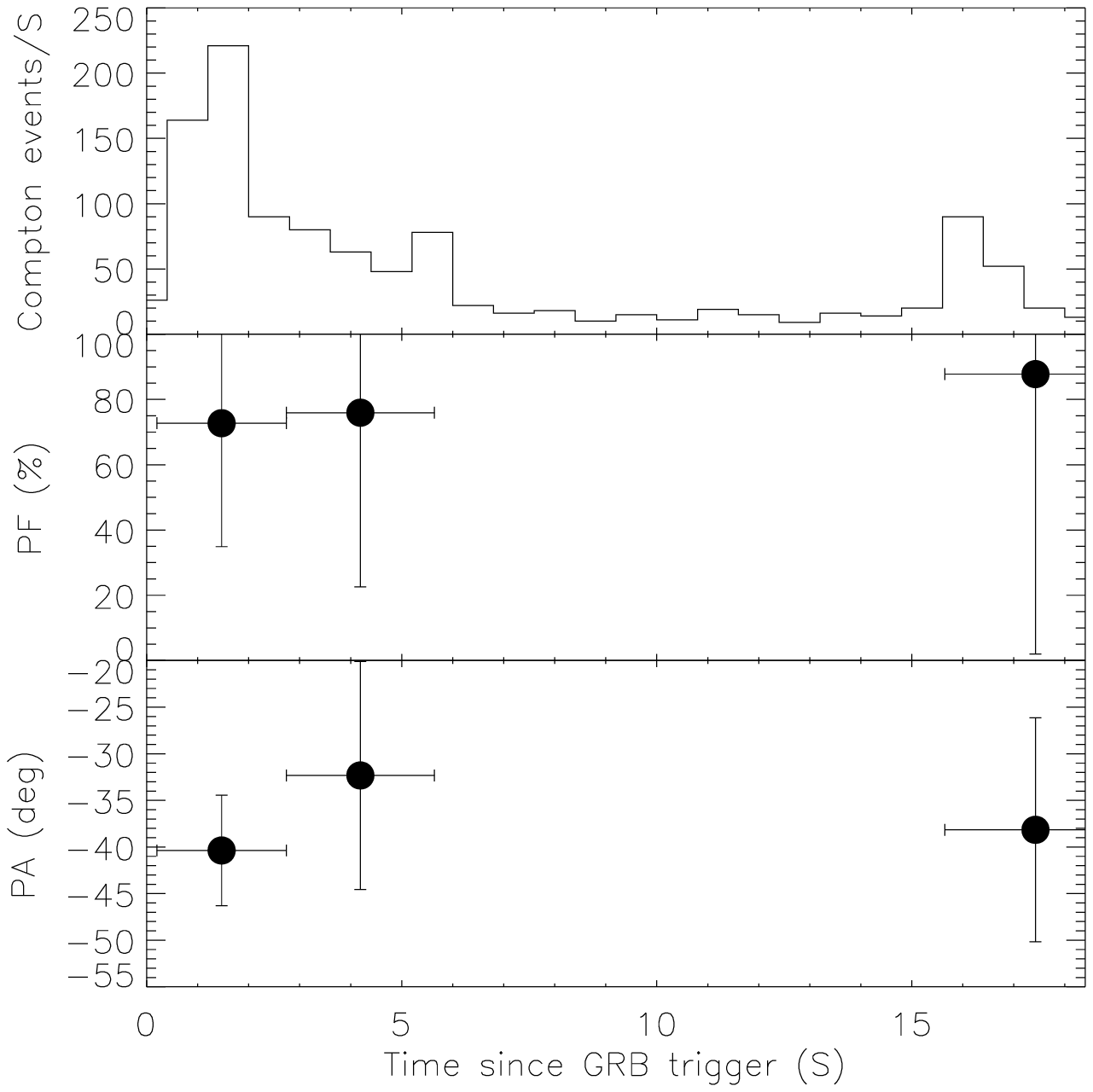}
\caption{The Compton double events (top panel), polarisation fraction (middle panel) and the polarisation angle (bottom panel) of
	GRB~160802A as a function of time (\S\ref{sec:polarization}).  } 
	\label{fig:160802A_polarisation}
\end{figure}

GRB~160802A shows two distinct peaks and, further, it  also shows two distinct
phases of spectra in the first peak: the first phase (covering 0 -- 2.5~s, see
Figure~\ref{fig:160802A_alpha_evolution}) having high $E_p$ and hard low
energy spectral index, and the second phase with more modest values of $E_p$ and
a softer low energy spectral index. Most of the time $E_p$ is above 200 keV. CZT
Imager is sensitive for polarisation in the 100 -- 400~keV range, the efficiency
peaking at lower energies. Hence, for this analysis we restrict ourselves to the
energy range of 110 -- 175~keV, consistently below the spectral peak ($E_p$).
To investigate the variation of the polarisation characteristics with the pulse
characteristics, we undertook polarisation measurements in three distinct time
intervals: (i) $\Delta t<$ 2.74 s (measured from the Fermi trigger time), 
(ii)   2.74~s $< \Delta t <$ 5.64~s, and 
(iii)  15.65~s $< \Delta t <$ 20.34~s (second pulse). 
The results are shown in Figure~\ref{fig:160802A_polarisation}. 
  
It is interesting to note that in spite of a large variation in the flux as well
as low energy spectral index, the polarisation value remains high throughout the
burst.
  
\section{Discussion and Conclusions}
\label{sec:conclusions_discussions}
We have carried out systematic analysis of spectral and polarisation data from
\Fermi and AstroSat/CZTI respectively for GRB~160802A.   Our analysis derives a
number of important constraints as follows.  
\begin{enumerate}[label=(\alph*)]
\item We obtain hard $\alpha$ values most of which lie above $-0.67$, the line
	of death of synchrotron emission in slow cooling limit
	\citep{Cohen:1997, Crider:1997, Ghirlanda:2003, Preece:1998, Goldstein:2013}, throughout
	the burst duration including the first and second episodes
	(0--7\,s and 12--20\,s). In coarse time bins though the spectra
	show a hint of a Black Body (BB) component, this is most
	probably an artefact of evolving peak energy
	\citep{Burgess:2015}. In the finer resolved time bins,
	the BB component is never statistically required and $\alpha$
	still remains above the line of death
	\citep{Crider:1997, Preece:1998, Ghirlanda:2003}.
\item During the initial part of the first pulse $\alpha$ is found to get harder
	with time, even reaching values  $> 0$. 
\item The peak energy $E_p$ in general shows a decreasing trend across each major
	episodes of emission. However, note that the first episode possesses
	multiple pulses in the light curve (Fig.~\ref{fig:160802A_LC}), in contrast
	to the second episode which is a smooth single pulse. Thereby,
	the $E_p$ evolution in the first episode is more complex, starting with
	a hard-to-soft (HTS) evolution, followed by an intensity
	tracking behaviour at later times.
\item Interestingly, it can also be noted that across the first episode (significantly in the first pulse of the episode), the
	time evolution of $E_{\rm peak}$ and $\alpha$ have a negative
	correlation in contrast to what is typically observed in bursts
	\citep{Kaneko:2006}, whereas a positive correlation can be
	observed in the second episode.
\item No high energy emission above 2 MeV is observed which suggests a cutoff
	above this energy. 
\item Finally, strong polarisation, $\pi = 85 \pm 29 \%$ in energy range
	100--175\,keV is observed, in both the episodes.
\end{enumerate}

On an average the $\beta$ of the Band function fit to the spectrum is
$-3.4$ with standard deviation $<\sigma> ~=~1.9$ for the burst (Table~\ref{tab:specfitting}). This indicates a
strong curvature at the high energy end of the spectrum. In such a scenario a
negative correlation between $E_p$ and $\alpha$ implies the following:
i) The decrease in $E_p$ with time is accompanied by a narrowing of the
spectrum. ii) As the $E_p$ values are well constrained and are located
well within the energy window, the inferred hardening of $\alpha$ is real, and
not an artefact caused by $E_p$ being close to the edge of the
observing band. iii) This behaviour also suggests that the overlapping of spectra in each time bin due 
to time integration is minimum. If significant overlapping of spectra with decreasing $E_p$ did occur, 
then it would result in an average spectrum with softer $\alpha$. Also note that in segments of evolution 
where $E_p$ is not found to be decreasing, the corresponding $\alpha$ values are relatively soft, 
indicating that such segments may consist of overlap of several pulses \cite{Ryde:1999}.
  
The obtained hard values of $\alpha > -0.67$ lead to the inference that the observed spectra are inconsistent
with the optically thin fast ($\alpha = -1.5$) and slow cooling ($\alpha =-0.67$) synchrotron emission models 
(\citealt{Katz:1994}, \citealt{Rees:1994}, \citealt{Tavani:1996},
\citealt{Sari:1998}) as well as with the fast cooling synchrotron emission model 
from a decaying magnetic field ($\alpha \leq -0.8$) \citep{Uhm:2014}.
A smooth Band function with a hard $\alpha$ may suggest a photospheric emission
scenario wherein continuous dissipation occurs at high optical depths extending
till the photosphere \citep{Beloborodov:2010}. If so, then the spectral peak
would be unpolarised as it corresponds to the Wien peak formed at higher optical
depths. However, during the interval 1.6~s to 5.64~s post trigger, the $E_p$
lies between 100--300~keV, within the CZTI band where high polarisation
is observed.  This observation is thus incongruous with the above model.
Polarisation imparted by scattering does not help either, since at large optical
depths multiple scatterings tend to wash out the directionality.  Therefore the
observed high polarisation cannot be expected from the subphotospheric
dissipation model based on Comptonisation (see \citealt{Lundman:2016} for more
details).

Another non-thermal process which could be compatible with the observed hard
values of $\alpha$ is Jitter radiation \citep{Medvedev:2000}, wherein 
small scale random magnetic turbulence (correlation length, $\lambda_B < $
Larmor radius of the electron) can result in deflection of the electrons on
scales less than the beaming angle. The emission thus produced can result in
hard spectra ($\alpha >-0.67$) and can in principle be as large as $+0.5$. The
Jitter emission spectrum, however, is better modeled by a sharp broken power
law instead of a smooth function such as Band function. Jitter radiation, in contrast 
to synchrotron emission, extends to frequencies well beyond the synchrotron critical 
frequency: up to $a^3 \omega_c$ where $a=R_L/\lambda$, $R_L$ the Larmor radius, $\lambda$ the 
characteristic scale of turbulence in the outflow, and $\omega_c$ is the characteristic 
synchrotron frequency. Since $a \gg 1$, emission extending to 
high energy gamma rays are expected (\citealt{Aharonian:2002}, \citealt{Kelner:2013}). However, here 
we find a cutoff in emission around 2 MeV.  High
polarisation in case of Jitter radiation can be achieved only in a specific
geometry where the magnetic field turbulence is constrained within a slab (or
plane) that is viewed nearly edge on.  This can result in polarisation degrees
as high as $90 \%$ \citep{Prosekin:2016}. However, it is important to note that
irrespective of the observing angle the degree of polarisation  is highest at
energies much beyond the spectral peak. Around the spectral peak, the
polarisation is expected to be relatively low with values $\le 40\%$ (see Fig 1
of \citealt{Prosekin:2016}). 

Hard $\alpha$ values also suggests that our line of
sight is not significantly off the burst axis \citep{Lundman:2013}, and that the
central engine is active throughout the burst duration. At large off-axis angles and when 
the central engine is off, the observed emission would be dominated by high latitude emissions, 
resulting in a softer spectrum both in terms of $E_p$ and $\alpha$ due to lower Doppler boost and superposition 
of spectra respectively. This would result in an average spectrum 
with region below the spectral peak $-1 \geq \alpha \leq -0.5$ \citep{Lundman:2013,Peer:2008}, softer 
than what is observed here.
Both the emission models discussed above, while being capable of generating spectra 
nearly compatible with those observed, find it difficult to explain the high degree of 
polarisation around the spectral peak. The observed high polarisation therefore results 
most likely from the viewing geometry, as envisaged by \cite{Waxman:2003}, who showed that 
bright and highly polarised emission can be seen when the observer's line of sight makes an 
angle $\theta_{\rm j} \leq \theta_{\rm v} \leq \theta_{\rm j} + 1/\Gamma$ from the jet axis, where $\theta_j$ is the jet opening angle, $\theta_v$ is the viewing angle with respect to the 
axis of the jet and $\Gamma$ is the Lorentz factor of the outflow. This 
also requires a strong asymmetry in observed emission within the off-axis viewing cone as could 
be obtained in a ``top-hat'' jet model, but not in ``structured jet'' models where emissivity 
drops slowly away from the jet axis. A sharp drop in emissivity beyond the edge of the jet is 
also suggested by the observed hard $\alpha$ values. In case of structured jets, the emission 
viewed off-axis would be dominated by that from high latitudes, resulting in a softer spectrum 
both in terms of $E_{\rm peak}$ and $\alpha$ due to lower Doppler boost and superposition of 
spectra respectively, contrary to what is observed. On the other hand, in case of a "top-hat" jet, hardly 
any high latitude emission is expected and the hard spectrum can survive even when observed close to the edge of the jet.
The hard $\alpha$ then suggests subphotospheric dissipation to be the underlying
emission mechanism. In this model Comptonisation can yield high polarisation
since orthogonal Thomson scattering in the rest frame dominates near the edge of
the jet. 
In contrast, regardless of the geometry, Jitter radiation cannot produce the observed high level of polarisation near the spectral peak.

In our current analysis, a spectral cutoff is observed beyond $2 \,{\rm MeV}$.
By using the argument of $\gamma - \gamma$ attenuation \citep{Lithwick:2001}, we
can constrain the lower limit of Lorentz factor of the outflow for the first and
second episodes  to be $\geq 78 \pm 23$ and $74 \pm 23$ respectively. This in
turn gives an upper limit on the beaming angle for the first episode (second
episode), $1/\Gamma \leq 0.73 \pm 0.22$ degrees ($0.77 \pm 0.24 $ degrees). This
value is consistent with the viewing angle geometry ($\theta_v/\theta_j$)
suggested by \cite{Waxman:2003}. Alternatively, under the assumption
of a narrow jet, the jet opening angle is inferred to be $\theta_j \sim 1/\Gamma
\sim 1^{\circ}$, similar to  the lowest $\theta_j$ deduced from jet breaks
observed in afterglows \citep{Racusin:2009}. 

Thus, combining the spectral analysis and the polarisation measurements, we deem
it the most likely that the observed emission from GRB~160802A is due to subphotospheric dissipation 
taking place within a narrow GRB jet viewed along its boundary, with jet emissivity dropping sharply away from its edge.

To summarize, the spectral and timing properties are rich sources of information about
emission mechanisms. Polarization of the prompt emission helped us to further narrow down 
the emission mechanisms. It also helped us to infer the jet geometry as high polarization is observed in case of this GRB.
However, by the combined constraints given by spectro-polarimetric properties we
need a narrow jet viewed along its edge. It is, however, on our radar to see in future for similar bright
bursts whether we obtain a high polarization always by geometric effects only. Observation
of afterglows is also a deciding factor as then we can measure the jet opening angles 
and even rule out such geometries when coupled with harder values of low energy spectral indices.

\begin{table*}
\begin{small}
\caption{Summary of spectral fitting for the two emission episodes in GRB~160802A}
\label{tab:spec_gbm_lat_3GeV}
\begin{center}
\begin{tabular}{p{3.0cm}p{3.0cm}p{3.0cm}p{3.0cm}p{3.0cm}}
\hline
Episode 1:& && & \\
($0 ~-~ 7$ s)& && &\\
\hline \hline
Parameters$^a$ & B                                         &BB+B                                  & CPL                               & BB+ CPL                                               \\ \hline \hline
$\alpha$ &  $-0.52_{-0.03}^{+0.04}$ & $-0.73_{-0.05}^{+0.06}$   &--                                 &--                                                 \\
$\beta$       & $-2.6_{-0.1}^{+0.1}$              &$-3.8_{-\infty}^{+0.7}$           &--                                 &--                                                    \\
$E_{\rm{p}}$ (keV) & $297_{-14}^{+15}$          & $429_{-36}^{+37}$       &--                                 & --                                                     \\\hline
$E_c$ (MeV)   &    --                                     &   --                                 & $247_{-12}^{+13}$      & $348_{-34}^{+41}$                        \\ \hline
$kT_{BB}$ (keV)  &   --                          &$35_{-4}^{+3}$        &  --                               &$35_{-3}^{+3}$    \\ \hline
$\Gamma_c$ &--                                 &--                                    &$0.60_{-0.03}^{+0.03}$  &$0.70_{-0.05}^{+0.05}$ \\ \hline
Flux$^b$ & $7.7$ &$8.0$                      & $7.9$ & $8.0$        \\
\hline \hline
pgstat/dof/BIC     &   395/341/430                 &367/339/414                  &  435/342/464                 & 368/340/409                       \\
$\Delta$(BIC)$^c$      &   --                                 &  16                           &                                  &   55                                               \\
\hline \hline
Episode 2:& && & \\
$15 ~-~ 20$ s & && &\\ 
\hline \hline
$\alpha$ &  $-0.78_{-0.11}^{+0.11}$ & $-1.1_{-0.1}^{+0.2}$   &--                                 &--                                                \\
$\beta$       & $-2.4_{-0.2}^{+0.1}$              &$-3.8_{-\infty}^{+3.0}$           &--                                 &--                                                   \\
$E_{\rm{p}}$ (keV) & $122_{-13}^{+17}$          & $267_{-86}^{+74}$       &--                                 & --                                                     \\\hline
$E_c$ (MeV)   &    --                                     &   --                                 & $173_{-20}^{+24}$      & $326_{-83}^{+139}$                         \\ \hline
$kT_{BB}$ (keV)  &   --                          &$17_{-4}^{+3}$        &  --                               &$17_{-3}^{+3}$     \\ \hline
$\Gamma_c$&--                                 &--                                    &$1.00_{-0.06}^{+0.06}$  &$1.2_{-0.1}^{+0.1}$ \\ \hline
Flux$^b$ & $2.3$                      & $2.4$ & $2.2$& $2.4$     \\
\hline \hline
pgstat/dof/BIC     &   399/341/434                   &393/339/440                 &  423/342/458                & 393/340/434                      \\
$\Delta$(BIC)$^c$      &   --                                 &  -6                         &                                  &  24                                          \\
\hline \hline
\multicolumn{5}{l}{$^a$Parameters are $\alpha$, $\beta$ and E$_p$ for Band (B) model, kT$_{BB}$ for the black body (BB) model, cut-off energy, E$_c$, and}\\
\multicolumn{5}{l}{index $\Gamma_c$ for the cut-off power law (CPL) model.}\\
\multicolumn{5}{l}{$^b$Flux in the units of  10$^{-6}$ erg cm$^{-2}$ in 8 - 1000 keV.}\\
\multicolumn{5}{l}{$^c$ $\Delta$BIC  is the decrease in BIC with respect to  the Band model or the CPL model.}\\
\end{tabular}
\end{center}
\end{small}
\end{table*}

\begin{table*}[h]
\begin{center}
\caption{Spectral fit to GBM data in the coarse time intervals described in \S\ref{sec:spectral_analysis_tr_coarse}.
Errors on the parameters are corresponding to a 90\% confidence region.}
\label{tab:coarse_results}
\footnotesize 
\begin{tabular}{p{1.25cm}p{1.5cm}cccccc}
\tableline\tableline
Interval & Model & $\alpha$ & $\beta$ & $E_{\rm p}$ & $kT_{\rm BB}$  & PGstat/dof/BIC & Prefered model \\
         &       &$\Gamma_1$&$\Gamma_2$ & $E_{break}$ &              &  &\\
\\
{[s]}  & & &  & [keV] & [keV] &  &   \\
\tableline
$i$ & B & $-0.42_{-0.05}^{+0.05}$  & $-3.45_{-\infty}^{+0.60}$ &$546_{-37}^{+38}$ & &    386/343/409 & \\
 $(0.0,~1.0)$ & BB+B  &$-0.51_{-0.04}^{+0.04}$   & $-3.0_{-3.0}^{+9.2}$ & $680_{-66}^{+46}$ & $47_{-9}^{+9}$ &  375/341/411 & Band/BB+Band\\
              & bknpower &$0.64_{-0.04}^{+0.04}$   & $2.31_{-0.09}^{+0.09}$ & $274_{-20}^{+21}$ &  & 469/343/492 &\\ \hline
$ii$ & B &   $-0.21_{-0.06}^{+0.06}$ &  $-2.76_{-0.28}^{+0.19}$  &  $307_{-18}^{+19}$ & --- &  360/345/383  & \\
$(1.0,~2.0)$ & BB+B &   $-0.40_{-0.12}^{+0.12}$ &   $-3.5_{-\infty}^{+0.6}$  &   $412_{-49}^{+61}$ & $37_{-7}^{+6}$ &  345/341/380 &Band/BB+Band \\
             & bknpower &   $0.57_{-0.04}^{+0.04}$ &   $2.32_{-0.07}^{+0.08}$  &   $178_{-10}^{+11}$ &  & 446/343/469 &\\ \hline
\hline
$iii$   & B&  $-0.31_{-0.09}^{+0.10}$ &  $-2.79_{-0.49}^{+0.27}$  &  $213_{-24}^{+18}$ & --- &  402/345/426 &\\
 $(2.0,~3.0)$& BB+B &   $-0.63_{-0.16}^{+0.23}$   &    $-4.34_{-\infty}^{+1.55}$  &   $304_{-67}^{+57}$ & $33_{-6}^{+5}$ &   398/341/433 & Band\\
             & bknpower &   $0.72_{-0.05}^{+0.05}$ &   $2.39_{-0.11}^{+0.12}$  &   $136_{-11}^{+11}$ &  & 426/343/450  &\\\hline

$iv$  & B &  $-0.6_{-0.1}^{+0.2}$ &  $-2.8_{-2.5}^{+0.4}$  &  $192_{-30}^{+26}$ & --- &  387/345/410 & \\
$(3.0, 4.0)$& BB + B &   $-0.81_{-0.21}^{+0.21}$ &   $-3.26_{-\infty}^{+0.85}$  &   $258_{-43}^{+90}$ & $26_{-7}^{+7}$ &   383/341/418  & Band/bknpower\\
             & bknpower &   $0.90_{-0.07}^{+0.07}$ &   $2.21_{-0.10}^{+0.12}$  &   $104_{-10}^{+12}$ &  & 341/343/414 &\\\hline

$v$& B &   $-0.76_{-0.20}^{+0.20}$ &   $-2.48_{-\infty}^{+0.28}$  &   $133_{-23}^{+37}$ &   &338/345/362 & \\
$(4.0, 5.0)$& BB + B &   $-1.05_{-0.29}^{+1.78}$&   $-2.94_{-\infty}^{+0.94}$  &   $190_{-123}^{+121}$ &  $21_{-21}^{+71}$  &  337/343/372 & Band \\
             & bknpower &   $1.12_{-0.17}^{+0.09}$ &   $2.32_{-0.25}^{+0.24}$  &   $94_{-26}^{+19}$ &  & 341/343/364 &\\\hline


$vi$& B &   $-0.76_{-0.08}^{+0.08}$ &   $-9.36_{-\infty}^{+6.64}$  &   $252_{-31}^{+25}$ & --- &   414/345/437  &\\
$(5.0, 6.0)$& BB + B &   $-0.95_{-0.20}^{+0.21}$   &    $-10_{-\infty}^{+0}$  &   $304_{-74}^{+109}$ & $36_{-36}^{+74}$ &   411/343/446  & Band\\
             & bknpower &   $1.00_{-0.06}^{+0.06}$ &   $2.44_{-0.16}^{+0.26}$  &   $151_{-17}^{+27}$ &  & 421/343/444 &\\\hline

$vii$& B &   $-0.17_{-0.80}^{+1.83}$ &   $-2.13_{-0.45}^{+0.23}$  &   $58_{-21}^{+35}$ & --- &    323/345/347 &\\
$(6.0, 7.0)$& BB + B &   $0.28_{-1.72}^{+\infty}$   &    $-1.90_{-0.54}^{+0.28}$  &   $38_{-21}^{+181}$ & $17_{-17}^{+\infty}$ & 323/343/358  & Band/bknpower\\
             & bknpower &   $0.72_{-0.58}^{+0.57}$ &   $2.06_{-0.18}^{+0.35}$  &   $36_{-8}^{+31}$ &  &324/343/348  &\\\hline

$viii$& B &   $-0.60_{-0.11}^{+0.12}$ &   $-2.22_{-0.13}^{+0.10}$  &   $143_{-16}^{+19}$ & --- &  389/345/413  &\\
$(15.0, 17.0)$& BB + B &   $-0.96_{-0.13}^{+0.14}$   &    $-2.74_{-\infty}^{+0.36}$  &   $280_{-57}^{+79}$ & $19_{-3}^{+3}$ &  375/343/410  & Band/BB+Band\\
             & bknpower &   $1.00_{-0.06}^{+0.05}$ &   $2.12_{-0.06}^{+0.07}$  &   $88_{-8}^{+8}$ &  & 410/343/432 &\\\hline
$ix$& B &   $-1.0_{-0.3}^{+0.7}$ &   $-2.4_{-0.6}^{+0.3}$  &   $65_{-23}^{+18}$& --- &   364/345/387  &\\
$(17.0, ~ 18.0)$& BB + B &   $1.2_{-1.4}^{+\infty}$   &    $-2.1_{-0.2}^{+0.1}$  &   $24_{-6}^{+22}$ & $20_{-5}^{+6}$ &   359/343/394 & Band/bknpower\\
             & bknpower &   $1.40_{-0.36}^{+0.15}$ &   $2.32_{-0.22}^{+0.25}$  &   $53_{-20}^{+23}$ &  & 394/343/391 &\\\hline
\tableline
\end{tabular}
\end{center}
\end{table*}

\begin{table*}
\caption{Detailed time-resolved Spectral fitting}
\label{tab:specfitting}
\begin{center}
\begin{tabular}{cc|cccccc}
\hline
\hline
Sr. no. &(t$_1$,t$_2$) &$kT_{BB}$& $\alpha$ & $\beta$ & E$_p$ (keV) & PGSTAT/dof/BIC& Prefered model \\
        &              &         &$\Gamma_1$&$\Gamma_2$ &  $E_{break}$ &  &\\
\hline
1&(-0.26, 0.32)&&$-0.55_{-0.09}^{+0.05}$& $-5.0_{-\infty}^{+2.0}$& $951_{-127}^{+168}$  &$ 343 / 343/ 366$ & Band/bknpower\\
&&$4.3_{-4.3}^{+3.0}$&$-0.40_{-0.15}^{+0.17}$& $-10_{-\infty}^{+20}$& $850_{-115}^{+9055}$  &$ 335 / 341/ 370$ &\\
&&&$0.84_{-0.05}^{+0.05}$&  $7_{-3}^{+\infty}$ &$1020_{-188}^{+135}$  &$ 341 / 343 /364$ &\\
2&(0.32, 0.48)&&$-0.41_{-0.08}^{+0.08}$&$-8.0_{-\infty}^{+5.0}$& $685_{-71}^{+84}$  &$325/  343/349 $ & Band\\
&&$48_{-48}^{+460}$&$-0.43_{-0.15}^{+0.14}$&$-10_{-\infty}^{+20}$ & $746_{-78}^{+167}$  &$324/  341/359  $ &\\
&&&$0.72_{-0.19}^{+0.06}$&$2.9_{-0.7}^{+0.6}$ & $504_{-212}^{+106}$  &$353/  343/376 $ & \\
3&(0.48, 0.64)&&$-0.36_{-0.12}^{+0.13}$&$-3.4_{-\infty}^{+0.8}$& $491_{-62}^{+74}$  &$357 /  343/381 $& Band\\
&&$47_{-20}^{+21}$&$-0.5_{-0.2}^{+0.2}$&$-9.4_{-\infty}^{+9.5}$& $586_{-123}^{+142}$  &$ 352/  341/ 388 $&\\
&&&$0.64_{-0.13}^{+0.08}$&$2.4_{-0.3}^{+0.3}$& $278_{-72}^{+45}$   &$374 /  343/ 398$&\\
4&(0.64, 0.83)&&$-0.20_{-0.15}^{+0.17}$&$-3.36_{-\infty}^{+0.80}$& $326_{-44}^{+48}$ &$318/   343/341$&Band\\
&&$25_{-11}^{+26}$&$-0.18_{-0.27}^{+0.54}$&$-6_{-\infty}^{+16}$& $372_{-58}^{+79}$ &$315/   341/350$&\\
&&&$0.61_{-0.18}^{+0.10}$&$2.5_{-0.4}^{+0.4}$& $213_{-58}^{+48}$ &$339/ 343/ 362$&\\
5&(0.83, 0.98)&&$-0.02_{-0.20}^{+0.22}$& $-2.34_{-0.45}^{+0.20}$& $283_{-41}^{+61}$  &$412/  343/435$&Band\\
&&$17_{-6}^{+7}$&$1.8_{-2.2}^{+\infty}$& $-2.3_{-1.1}^{+0.2}$& $245_{-40}^{+166}$  &$408/  341/443$&\\
&&&$0.50_{-0.11}^{+0.10}$&$2.2_{-0.2}^{+0.2}$ & $181_{-25}^{+23}$   &$421/  343/ 444$&\\
6&(0.98, 1.1)&&$-0.03_{-0.18}^{+0.20}$& $-2.36_{-0.33}^{+0.20}$& $316_{-46}^{+55}$ &$368/  343/392$&Band\\
&&$18_{-5}^{+23}$&$0.5_{-1.0}^{+4.0}$& $-2.4_{-1.1}^{+0.2}$& $314_{-78}^{+450}$ &$365/  341/400$&\\
&&&$0.40_{-0.12}^{+0.11}$&$2.2_{-0.1}^{+0.2}$ & $181_{-25}^{+30}$ &$382/  343/405$&\\
7&(1.1, 1.2)&&$-0.16_{-0.21}^{+0.48}$& $-2.82_{-1.06}^{+0.38}$& $372_{-55}^{+90}$   &$310 /  343/333$&Band\\
&&$8_{-8}^{+\infty}$&$-0.15_{-0.60}^{+0.50}$& $-2.8_{-0.9}^{+0.4}$& $361_{-55}^{+322}$   &$ 310/  341/345$&\\
&&&$0.6_{-0.1}^{+0.1}$& $2.4_{-0.2}^{+0.2}$ & $226_{-35}^{+38}$  &$321/  343/ 344$&\\
8&(1.2, 1.3)&&$-0.11_{-0.17}^{+0.20}$& $-2.63_{-0.80}^{+0.32}$& $290_{-42}^{+48}$ &$333/  343/ 356$&Band\\
&&$29_{-14}^{+13}$&$-0.2_{-0.3}^{+0.7}$& $-3.3_{-\infty}^{+0.9}$& $401_{-120}^{+125}$ &$ 329/  341/364$&\\
&&&$0.4_{-0.1}^{+0.2}$& $2.2_{-0.1}^{+0.2}$& $145_{-16}^{+42}$ &$352/  343/ 375$&\\
9&(1.3, 1.5)&&$-0.20_{-0.13}^{+0.17}$& $-3.3_{-\infty}^{+0.7}$& $345_{-48}^{+39}$    &$364 /  343/388$&Band\\
&&$37_{-12}^{+13}$&$-0.30_{-0.23}^{+0.24}$& $-3.6_{-\infty}^{+1.0}$& $427_{-78}^{+126}$    &$360 /  341/395$&\\
&&&$0.5_{-0.1}^{+0.1}$& $2.3_{-0.2}^{+0.2}$&  $175_{-23}^{+27}$    &$379 /  343/ 402$&\\
10&(1.5, 1.6)&&$-0.23_{-0.13}^{+0.18}$& $-3.4_{-\infty}^{+0.9}$& $364_{-56}^{+45}$  &$315/ 343/338$&Band\\
&&$38_{-16}^{+15}$&$-0.34_{-0.23}^{+0.26}$& $-10_{-\infty}^{+0}$& $455_{-75}^{+128}$  &$311/ 341/346$&\\
&&&$0.50_{-0.12}^{+0.15}$& $2.3_{-0.2}^{+0.3}$& $183_{-25}^{+55}$   &$333/ 343/356 $&\\
11&(1.6, 1.72)&&$-0.36_{-0.19}^{+0.18}$& $-3.2_{-\infty}^{+0.8}$& $288_{-47}^{+50}$  &$309/   343/332$&Band\\
&&$13_{-8}^{+37}$&$-0.1_{-0.6}^{+1.6}$& $-3.2_{-\infty}^{+0.8}$& $288_{-76}^{+117}$  &$ 306/   341/341$&\\
&&&$0.72_{-0.09}^{+0.08}$& $2.5_{-0.2}^{+0.3}$ & $186_{-30}^{+26}$   &$321/   343/344$&\\
12&(1.72, 1.9)&&$-0.3_{-0.2}^{+0.2}$& $-3.5_{-\infty}^{+0.9}$& $250_{-34}^{+37}$  &$348/  343/ 371$&Band\\
&&$41_{-6}^{+8}$&$-0.5_{-0.3}^{+0.5}$& $-10_{-\infty}^{+0}$& $306_{-49}^{+93}$  &$346/  341/381$&\\
&&&$0.68_{-0.15}^{+0.11}$& $2.5_{-0.3}^{+0.4}$ & $157_{-31}^{+36}$   &$355/  343/ 378$&\\
13&(1.9, 2.05)&&$-0.12_{-0.18}^{+0.21}$& $-2.8_{-1.2}^{+0.4}$& $236_{-32}^{+35}$ &$337 /  343/360 $&Band/bknpower\\
&&$41_{-7}^{+7}$&$-0.7_{-0.3}^{+0.3}$& $-10_{-\infty}^{+0}$& $438_{-122}^{+241}$  &$333 /  341/368 $&\\
&&&$0.5_{-0.1}^{+0.1}$& $2.4_{-0.2}^{+0.2}$ &$142_{-15}^{+19}$  &$336/  343/ 359$&\\
14&(2.05, 2.2)&&$-0.06_{-0.20}^{+0.20}$& $-2.4_{-0.4}^{+0.2}$& $216_{-32}^{+43}$ &$319/  343/ 342$&Band\\
&&$32_{-8}^{+7}$&$-0.5_{-0.2}^{+0.3}$& $-3.5_{-\infty}^{+1.1}$& $402_{-124}^{+149}$ &$315/  341/350$&\\
&&&$0.52_{-0.14}^{+0.12}$& $2.2_{-0.1}^{+0.2}$ &$129_{-19}^{+24}$  &$326/  343/ 350$&\\
15&(2.2, 2.4)&&$-0.37_{-0.15}^{+0.18}$& $-3.0_{-\infty}^{+0.5}$& $225_{-30}^{+34}$   &$332/  343/355$&Band\\
&&$60_{-60}^{+\infty}$&$-0.1_{-0.9}^{+4.7}$& $-2.1_{-1.6}^{+0.3}$& $95_{-41}^{+211}$   &$331/  341/366$\\
&&&$0.82_{-0.12}^{+0.08}$&$2.7_{-0.4}^{+0.4}$ & $173_{-37}^{+26}$    &$338/  343/ 361$&\\
16&(2.4, 2.74)&&$-0.2_{-0.2}^{+0.2}$& $-2.8_{-\infty}^{+0.4}$& $181_{-26}^{+34}$  &$413/ 343/436 $&Band\\
&&$27_{-22}^{+\infty}$&$-0.50_{-0.37}^{+2.42}$& $-9.4_{-\infty}^{+7.1}$& $246_{-87}^{+88}$  &$412/ 341/447 $&\\
&&&$0.64_{-0.12}^{+0.13}$& $2.3_{-0.2}^{+0.3}$  &$109_{-12}^{+27}$  &$421/ 343/444 $&\\
17&(2.74, 3)&&$-0.51_{-0.16}^{+0.22}$& $-3.40_{-\infty}^{+1.0}$& $214_{-40}^{+36}$   &$356/ 343/ 380$&Band\\
&&$31_{-20}^{+\infty}$&$-0.8_{-0.3}^{+0.2}$& $-10_{-\infty}^{+0}$& $279_{-27}^{+146}$   &$354/ 341/389$&\\
&&&$0.9_{-0.1}^{+0.1}$& $2.5_{-0.3}^{+0.4}$ &$142_{-26}^{+32}$   &$362/ 343/ 386$&\\ \hline
\end{tabular}
\end{center}
\end{table*}

\addtocounter{table}{-1}

\begin{table*}
\caption{Detailed time-resolved Spectral fitting (continued)}
\label{tab:specfitting}
\begin{center}
\begin{tabular}{cc|cccccc}
\hline
\hline
Sr. no. &(t$_1$,t$_2$) &$kT_{BB}$& $\alpha$ & $\beta$ & E$_p$ (keV) & PGSTAT/dof/BIC & Preferred model\\
        &              &         &$\Gamma_1$&$\Gamma_2$  & $E_{break}$  & & \\
\hline
18&(3, 3.45)&&$-0.60_{-0.20}^{+0.24}$& $-2.5_{-\infty}^{+0.3}$& $155_{-30}^{+41}$&$379/  343/402$&Band/bknpower\\
&&$27_{-16}^{+\infty}$&$-1.0_{-0.4}^{+0.6}$& $-2.6_{-\infty}^{+0.7}$& $231_{-166}^{+309}$&$378/  341/413$&\\
&&&$1.0_{-0.1}^{+0.1}$& $2.3_{-0.2}^{+0.2}$ & $103_{-16}^{+20}$ &$378/ 343/ 401$&\\
19&(3.45, 3.86)&&$-0.51_{-0.14}^{+0.15}$& $-10_{-\infty}^{+20}$& $204_{-20}^{+21}$&$407 / 343/430$&Band\\
&&$21_{-21}^{+\infty}$&$-0.55_{-0.30}^{+0.30}$& $-10_{-\infty}^{+0}$& $224_{-36}^{+70}$&$405 / 341/441$&\\
&&&$0.80_{-0.13}^{+0.14}$& $2.3_{-0.2}^{+0.4}$ & $107_{-15}^{+38}$    &  $418/343/441$& \\
20&(3.86, 4.2)&&$-0.6_{-0.1}^{+0.2}$& $-2.8_{-\infty}^{+0.5}$& $204_{-36}^{+38}$   &  $321/343/345$ &Band/bknpower\\
&&$29_{-16}^{+\infty}$&$-0.8_{-0.5}^{+0.8}$& $-2.8_{-\infty}^{+0.9}$& $235_{-171}^{+337}$   &  $321/341/356$&  \\
&&&$0.9_{-0.1}^{+0.1}$& $2.24_{-0.18}^{+0.29}$ & $115_{-20}^{+32}$     &$325/343/348 $&\\
21&(4.2, 5)&&$-0.75_{-0.22}^{+0.26}$& $-2.38_{-0.64}^{+0.25}$& $111_{-30}^{+28}$    &$368/343/391$&Band/bknpower\\
&&$69_{-69}^{+408}$&$3.7_{-4.7}^{+\infty}$& $-2.2_{-0.3}^{+0.2}$& $75_{-9}^{+75}$    &$367/341/402$&\\
&&&$1.1_{-0.1}^{+0.1}$& $2.2_{-0.1}^{+0.3}$ &$72_{-10}^{+25}$ &  $368/343/ 391$& \\
22&(5, 5.4)&&$-0.6_{-0.1}^{+0.1}$& $-9.2_{-\infty}^{+\infty}$& $246_{-19}^{+35}$  &  $366/343/390$ &Band/bknpower\\
&&$39_{-9}^{+\infty}$&$-1.0_{-0.3}^{+0.7}$& $-10_{-\infty}^{+0}$& $338_{-107}^{+186}$  &  $364/341/399$ &\\
&&&$0.91_{-0.09}^{+0.08}$& $2.5_{-0.2}^{+0.3}$ & $152_{-21}^{+26}$       &$367/343/ 391$&\\
23&(5.4, 5.6)&&$-0.60_{-0.12}^{+0.10}$& $-3.5_{-\infty}^{+0.9}$& $277_{-35}^{+43}$      &$369/343/392$  &Band\\
&&$27_{-27}^{+\infty}$&$-0.58_{-0.32}^{+0.35}$& $-3.61_{-\infty}^{+0.9}$& $282_{-46}^{+123}$      &$369/341/404$&\\
&&&$0.90_{-0.08}^{+0.08}$&$2.6_{-0.3}^{+0.3}$  &$184_{-31}^{+30}$   &$377/ 343/ 401$&\\
24&(15.7, 15.9)&&$-0.40_{-0.17}^{+0.24}$& $-2.3_{-0.5}^{+0.3}$& $247_{-54}^{+56}$       &$340/343/363$ &Band\\
&&$22_{-8}^{+10}$&$-0.48_{-0.30}^{+0.47}$& $-2.6_{-\infty}^{+0.4}$& $347_{-100}^{+179}$       &$335/341/370$&\\
&&&$0.62_{-0.14}^{+0.20}$&$2.0_{-0.1}^{+0.3}$ & $109_{-15}^{+55}$ &$351/ 343/374 $\\
25&(15.9, 16.0)&&$-0.48_{-0.17}^{+0.25}$& $-2.9_{-\infty}^{+0.5}$& $191_{-38}^{+36}$       &$337/343/360$ &Band/bknpower\\
&&$27_{-8}^{+7}$&$-0.8_{-0.3}^{+0.3}$& $-4.4_{-\infty}^{+2.0}$& $289_{-91}^{+136}$       &$333/341/369$&\\
&&&$0.76_{-0.11}^{+0.11}$&$2.3_{-0.1}^{+0.2} $  & $104_{-11}^{+15} $   &  $337/343/360$ \\
26&(16.0, 16.15)&&$-0.36_{-0.26}^{+0.40}$& $-2.40_{-0.51}^{+0.25}$& $127_{-29}^{+32}$   &$338/343/362$& Band/bknpower\\
&&$48_{-12}^{+81}$&$-0.07_{-0.90}^{+0.50}$& $-2.2_{-0.4}^{+0.2}$& $86_{-22}^{+191}$   &$336/341/371$& \\
&&&$0.76_{-0.15}^{+0.13}$& $2.18_{-0.12}^{+0.14}$ & $75_{-10}^{+11}$     &$340/343/363 $&\\
27&(16.15, 16.3)&&$-0.2_{-0.3}^{+0.4}$& $-2.4_{-0.3}^{+0.2}$& $102_{-20}^{+20}$ &  $332/343/356$ & Band\\
&&$15_{-3.3}^{+4.3}$&$-0.7_{-0.3}^{+0.1}$& $-10_{-\infty}^{+0}$& $214_{-30}^{+15}$ &  $329/341/364$ & \\
&&&$1.1_{-0.1}^{+0.1}$& $2.2_{-0.1}^{+0.3}$ & $72_{-10}^{+25}$   &  $368/343/ 365$ &\\
28&(16.3, 16.5)&&$-0.6_{-0.2}^{+0.3}$& $-2.5_{-0.5}^{+0.2}$& $99_{-16}^{+21}$   & $281/343/305$   & Band/bknpower  \\
&&$78_{-51}^{+\infty}$&$-0.5_{-0.2}^{+0.4}$& $-3.4_{-\infty}^{+1.1}$& $82_{-18}^{+25}$   & $281/343/316$ & \\
&&&$1.15_{-0.14}^{+0.11}$& $2.5_{-0.2}^{+0.3}$ & $86_{-18}^{+17}$       &$286/343/310$&\\
29&(16.5, 16.9)&&$-0.6_{-0.2}^{+0.3}$& $-2.5_{-0.4}^{+0.3}$& $86_{-13}^{+14}$    &   $356/343/379$& Band/bknpower\\
&&$8.5_{-8.5}^{+\infty}$&$-0.6_{-0.8}^{+1.8}$& $-2.6_{-\infty}^{+0.5}$& $106_{-23}^{+137}$    &   $355/341/390$&\\
&&&$1.15_{-0.14}^{+0.11}$& $2.5_{-0.2}^{+0.3}$ & $86_{-18}^{+17}$       &$286/343/380$&\\
30&(16.9, 17.6)&&$-1.0_{-0.2}^{+0.3}$& $-2.6_{-1.4}^{+0.3}$& $72_{-14}^{+17}$   &  $367/343/390$&Band/bknpower \\
&&$5_{-5}^{+1.7}$&$0.9_{-1.9}^{+\infty}$& $-2.5_{-0.4}^{+0.2}$& $82_{-13}^{+17}$   &  $365/341/400$ &\\
&&&$1.15_{-0.14}^{+0.11}$&$2.5_{-0.2}^{+0.3}$  & $86_{-18}^{+17}$      &$286/343/394$&\\
31&(17.6, 20.34)&&$0.05_{-1.01}^{+2.30}$& $-2.4_{-0.6}^{+0.3}$& $36_{-11}^{+15}$    &   $276/343/299$& bknpower\\
&&$5_{-5}^{+6}$&$-0.5_{-0.8}^{+5.0}$& $-2.7_{-1.4}^{+0.3}$& $78_{-17}^{+24}$    &   $273/341/308$&\\ 
&&&$1.15_{-0.14}^{+0.11}$& $2.5_{-0.2}^{+0.3}$ & $86_{-18}^{+17}$      &$286/343/302$&\\\hline
\end{tabular}
\end{center}
\end{table*}

\section*{Acknowledgements}
This research has
made  use  of  data  obtained  through  the  HEASARC  Online  Service,  provided  by  the  NASA/GSFC,  in  support  of  NASA  High
Energy  Astrophysics  Programs. This publication also uses the data from the AstroSat mission 
of  the Indian Space Research Organisation (ISRO), archived at the Indian Space Science Data Centre 
(ISSDC). CZT-Imager is built by a consortium of Institutes across India including Tata Institute of 
Fundamental Research, Mumbai, Vikram Sarabhai Space Centre, Thiruvananthapuram, ISRO Satellite Centre, 
Bengaluru, Inter University Centre for Astronomy and Astrophysics, Pune, Physical Research 
Laboratory, Ahmedabad, Space Application Centre, Ahmedabad: contributions from the vast technical team 
from all these institutes are gratefully acknowledged. 

\bibliography{GRB160802A_ref} 
\bibliographystyle{aasjournal}
\end{document}